\documentclass[usenatbib]{mn2e}

\usepackage{epsfig,latexsym,graphicx,lscape,amssymb}
\usepackage{marvosym}  
\usepackage{multicol}
\usepackage{comment}
\usepackage{pdflscape}
\usepackage{color, colortbl}
\usepackage{caption}
\usepackage{subcaption}
\newcommand\apj{ApJ}%
\newcommand\apjl{ApJ}%
\newcommand\apjs{ApJS}%
\newcommand\aap{A\&A}%
\newcommand\aapr{A\&A~Rev.}%
\newcommand\mnras{MNRAS}%
\newcommand\pasp{PASP}%
\newcommand\pasj{PASJ}%
\newcommand\nat{Nature}%
\newcommand\iaucirc{IAU~Circ.}%
\newcommand\araa{ARA\&A}%
\newcommand\ssr{Space Sci. Rev.}%

\definecolor{Gray}{gray}{0.9}

\begin{document}

\title[]{Broad iron line in the fast spinning neutron-star system 4U 1636--53}
\author[A. Sanna et al.]
       {Andrea Sanna$^{1}$\thanks{E-mail: sanna@astro.rug.nl}, Beike Hiemstra$^{1}$, Mariano M\'{e}ndez$^{1}$, Diego Altamirano$^{2}$, \newauthor Tomaso Belloni$^{3}$, and Manuel Linares$^{4}$\\
        $^{1}$Kapteyn Astronomical Institute, University of Groningen, P.O. Box 800, 9700 AV Groningen, The Netherlands\\
        $^{2}$Astronomical Institute Anton Pannekoek, University of Amsterdam, Science Park 904, 1098 XH Amsterdam, The Netherlands\\
        $^{3}$INAF-Osservatorio Astronomico di Brera, Via E. Bianchi 46, I-23807 Merate (LC), Italy\\
        $^{4}$Instituto de Astrof'sica de Canarias, V\'ia L\'actea,  E-38205 La Laguna, Tenerife, Spain}
\date{Accepted 2013 March 25. Received 2013 March 18; in original form 2011 November 7
      }

\pagerange{\pageref{firstpage}--\pageref{lastpage}}
\pubyear{2011}

\maketitle
\label{firstpage}

\begin{abstract}
We analysed the X-ray spectra of six observations, simultaneously taken with \textit{XMM-Newton} and \textit{Rossi X-ray Timing Explorer (RXTE)}, of the neutron star low-mass X-ray binary 4U 1636--53. The observations cover several states of the source, and therefore a large range of inferred mass accretion rate. These six observations show a broad emission line in the spectrum at around 6.5 keV, likely due to iron. We fitted this line with a set of phenomenological models of a relativistically broadened line, plus a model that accounts for relativistically smeared and ionised reflection from the accretion disc. The latter model includes the incident emission from both the neutron-star surface or boundary layer and the corona that is responsible for the high-energy emission in these systems. From the fits with the reflection model we found that in four out of the six observations the main contribution to the reflected spectrum comes from the neutron-star surface or boundary layer, whereas in the other two observations the main contribution to the reflected spectrum comes from the corona. We found that the relative contribution of these two components is not correlated to the state of the source. From the phenomenological models we found that the iron line profile is better described by a symmetric, albeit broad, profile. The width of the line cannot be explained only by Compton broadening, and we therefore explored the case of relativistic broadening. We further found that the direct emission from the disc, boundary layer, and corona generally evolved in a manner consistent with the standard accretion disc model, with the disc and boundary layer becoming hotter and the disc moving inwards as the source changed from the hard in to the soft state. The iron line, however, did not appear to follow the same trend. 

\end{abstract}


\section{Introduction}\label{sec:intro}
The X-ray spectrum of low-mass X-ray binaries (LMXBs) shows evidence of an accretion disc extending down close to the central object \citep[see e.g. review by][]{Done07}. The disc is assumed to be optically thick and geometrically thin, and is characterised by a thermal spectrum with typical temperatures of 0.2--1 keV \citep{Shakura73}. The thermal disc photons are Compton up-scattered in a hot coronal gas surrounding the disc and compact object, giving rise to the hard component usually seen in the spectrum \citep[e.g.][and references therein]{Barret00}. However, a fraction of these Comptonized photons is backscattered, and they irradiate the accretion disc. After being absorbed, reprocessed and re-emitted, these photons produce a reflection spectrum consisting of a continuum plus a complex line spectrum \citep[see review by][]{Fabian10}. Due to the high fluorescence yield and abundance, an important feature of the reflection spectrum is the Fe K$\alpha$ line at $\sim$6--7 keV. As a result of disc rotation and relativistic effects near the compact object, the reflection spectrum is blurred, with the Fe line being asymmetrically broadened \citep{Fabian89}. As the disc moves close to the central compact object, the relativistic effects become stronger, and consequently, the red wing of the Fe line -- mainly set by the gravitational redshift at the inner edge of the disc -- extends to lower energies. Hence, modelling of the iron line profile allows us to measure the size of the inner disc radius, which determines other spectral and timing properties of the source (e.g., disc temperature, characteristic frequency of the variability). If the compact object is a black hole (BH), modelling of the Fe line profile in the disc-dominated spectral state provides an estimate of the BH spin \citep[see e.g.][]{Miller07,Reynolds08}, assuming that the disc is truncated at the innermost stable circular orbit (ISCO). In case of a neutron star (NS), a measurement of the inner edge of the disc sets directly an upper limit on the NS radius, as the inner disc radius has to be larger than, or equal to, the equatorial radius of the star \citep[see e.g.][]{Bhattacharyya07,Cackett08}.

Since the first detection of a broad and asymmetric Fe emission line in the Seyfert-1 galaxy MCG--6--30--15 \citep{Tanaka95}, many similar lines have been observed in BH systems, both in Active Galactic Nuclei (AGN) and in X-ray binaries \citep[see reviews by][]{Fabian00,Miller07}. In all cases the line appears to be produced very close to the BH. Broad Fe lines are also detected in NS systems although they are weaker than in BH systems. With instruments like \textit{Chandra}, \textit{XMM-Newton} and \textit{Suzaku}, relativistically broadened Fe emission lines have been observed in many NS LMXBs (e.g., \citealt{Piraino07,Cackett08,Pandel08,diSalvo09,Cackett10}; see also \citealt{Ng10} for an almost complete list of NS LMXBs with Fe lines detected with \textit{XMM-Newton}).

Relativistically broadened Fe lines in NS spectra are usually fit using a phenomenological model for either a Schwarzschild or (maximum) Kerr metric, where the underlying continuum is modelled with a variety of direct emission components \citep[see e.g.][]{Bhattacharyya07, Lin07,Pandel08,Cackett08}. However, as the Fe emission line is a reflection signature, it should be accompanied with a reflection continuum extending over a wide bandpass; at the same time, the line and its reflection continuum should self-consistently determine the ionisation balance of the disc and include Compton scattering \citep[e.g.][]{George91,Ross93,Ross99}. An inappropriate modelling of the underlying continuum, as well as not including any of these processes in the analysis, could lead to incorrect results from the line modelling. Therefore, a self-consistent ionised reflection model should be used to study the continuum and Fe emission line in the spectra of LMXBs. (Note that, depending on the quality of the data, reflection and phenomenological models could describe the data equally well, even leading to reasonable results for the phenomenological model; see e.g. \citealt{Cackett10}.) Furthermore, the shape of the reflection spectrum not only depends on the ionisation state of the surface layers of the disc, but also on the spectral shape of the emission illuminating the accretion disc. Comptonized emission from the (disc) corona is the commonly assumed source of irradiation, but in the case of an accreting NS, the emission from the NS surface/boundary layer, can be as well irradiating the accretion disc. \citet{Cackett10} have studied a number of broad Fe emission lines in NS LMXBs where they used a blurred reflection model in which the blackbody component -- used to mimic the emission from the boundary layer -- was assumed to be the incident emission. \citet{Cackett10} found that this model fits the data well in most cases, supporting the idea that the boundary layer is indeed illuminating the disc \citep[see also][]{DAi10}. In our analysis (below), we test whether a disc corona or boundary layer was the most likely source of irradiation. We note that relativistic (plus Compton) broadening is not the only physical explanation for the width and the profile of iron lines in compact objects. For instance, \citet{Titarchuk03} proposed that asymmetric line profiles could originate from an optically thick flow launched from the disc, which expands or contracts at relativistic speeds. We will not explore this possibility in the paper, but refer the reader to the discussion in \citet{Ng10}. 
 
We study the Fe emission line and the reflection spectrum in six observations of the NS-LMXB 4U 1636--53, taken with the \textit{XMM-Newton} and the \textit{RXTE} satellites simultaneously. Besides the direct emission components, we fit these data using both phenomenological models and a relativistically-smeared ionised reflection model in which we investigated different sources of irradiation. In what follows, we first introduce the source 4U 1636--53, focusing on quantities relevant for this paper. In Section~\ref{sec:obs} we give details on the data used, and the procedure we followed to reduce them. In Section~\ref{sec:analysis} we describe the models, and explain the specific aspects of the spectral analysis. Here we also summarise the most important findings. We discuss our results and their implication on the standard accretion disc model in Section~\ref{sec:discussion}. There we also address the impact of the NS surface/boundary layer, on the reflection spectrum for the different spectral states. Furthermore, in that section we also discuss the results from the phenomenological line models. In Section~\ref{sec:summary} we summarise our results.
\subsection{4U 1636--53}\label{sec:1636}
Since its discovery in 1974 \citep{Willmore74,Giacconi74}, 4U 1636--53 has been observed over a wide range of wavelengths. Photometry of the optical counterpart (V801 Ara) revealed a short orbital period of $\sim$3.8 hr and a companion star with a mass of $\sim$0.4 M$_{\odot}$ \citep{Paradijs90,Giles02}. X-ray studies of 4U 1636--53 have shown a variety of rapid time variability, including \textit{i)} thermonuclear X-ray bursts \citep[e.g.][]{Hoffman77,Galloway06,Zhang11}, confirming the presence of a NS as the compact object, \textit{ii)} millisecond oscillations during the X-ray bursts, indicating a spin frequency of $\sim$581 Hz \citep{Zhang97,Giles02,Strohmayer02}, and \textit{iii)} kHz quasi-periodic oscillations \citep[kHz QPOs, e.g.][]{Wijnands97,Belloni07,Altamirano08,Sanna12}. From Eddington limited X-ray bursts, assuming a NS mass of 1.4 $M_{\odot}$ and a stellar radius of 10 km, \citet{Galloway06} estimated the distance to 4U 1636--53 to be $6.0\pm0.5$ kpc. Combining burst oscillations with phase-resolved optical spectroscopy of 4U 1636--53, \citet{Casares06} estimated the mass function and mass ratio to be $f(M)=0.76\pm0.47 M_{\odot}$ and $q=$ 0.21--0.34, respectively, where $q=M_2/M_1$ with $M_2$ the mass of the donor and $M_1$ the mass of the NS. \citet{Casares06} also showed that for a 0.48 $M_{\odot}$ donor, the NS mass is 1.6--1.9 $M_{\odot}$ and the inclination $i \simeq$ 60--71$^{\circ}$. The latter result is in conflict with the model of \citet{Frank87} as the lack of X-ray dips in the light curve of 4U 1636--53 sets an upper limit on the inclination of $i \leq 60^{\circ}$. However, in this model the derived upper limit for the inclination depends on the location where the dips are formed, and since the exact geometry is not known the minimum inclination to observe dips could be slightly larger. On the other hand, the inclination could not be larger than $\sim$75$^{\circ}$ since this would probably cause eclipses, which are not seen in this source.

4U 1636--53 is a persistent X-ray source, although it shows variations in intensity up to a factor of 10, following a narrow track in the colour-colour diagram (CD) and hardness-intensity diagram \citep[HID;][]{Belloni07,Altamirano08} with a $\sim$40-d cycle~\citep{Shih05,Belloni07}. The transition through the CD (or HID) is thought to be driven by changes in the mass accretion rate. The two distinct spectral states (hard, at the top right of the CD, and soft, at the bottom of the CD) are attributed to a different accretion flow configuration. In the soft state, where the X-ray intensity is high, the disc is hot and ionised, with an inner radius extending down to the ISCO (or down to the stellar surface in case the ISCO is within the NS, see e.g. \citealt{Done07}). As the intensity gradually decreases, the mass accretion rate eventually becomes too low to ionise the full accretion disc. The outer disc regions cool down, the viscosity decreases, and the overall mass accretion rate is reduced. Eventually, mass accretion rate at the inner regions of the disc should decrease, leading to a receding inner edge and the X-ray intensity reaches its minimum \citep{Done07}. The inner regions are likely replaced by a hot corona, and the spectrum hardens, which is the characteristic that defines the hard state. In the mean time, the secondary keeps transferring mass to the disc causing the density and temperature to increase, hence enhancing the viscous stress, and the mass accretion rate increases again. Eventually the disc gets ionised, the source rebrightens and makes a transition to the soft state. \citet{Shih05} have shown that this re-brightening cycle for 4U 1636--53 of $\sim$40 days is consistent with the viscous time-scale in the outer disc.

High signal-to-noise and moderate-resolution spectra of 4U 1636--53 revealed broad, asymmetric Fe emission lines. \citet{Pandel08} analysed three \textit{XMM-Newton/RXTE} observations, once when the source was in the transitional state (between the hard and the soft state), and twice when it was in the soft state. In all three spectra they found that the Fe line profile is consistent with a relativistically broadened line coming from the inner disc, which appeared to be at the ISCO for a non-rotating NS. This finding is partly contested by \citet{Cackett10} who fit the same spectra with a blurred reflection model that includes effects like Compton broadening and line emission from relevant elements. \citet{Cackett10} showed that fitting the spectrum with a self-consistent reflection model resulted in slightly larger values for the inner disc radius than when the spectrum was fit with the phenomenological \textsc{diskline} model \citep{Fabian89} used by \citet{Pandel08}. Nonetheless, both the results of \citet{Pandel08} and \citet{Cackett10} are based on data which were not corrected for pileup effects. As demonstrated by \citet{Ng10}, re-analyses of the same \textit{XMM-Newton} data considering pileup and background effects, suggested a different iron line profile. \citet{Ng10} found that the Fe lines in the three spectra of 4U 1636--53 used by \citet{Pandel08} appeared to be symmetric and could be well fit with a Gaussian profile, although, due to the statistics, a relativistic origin of the line could not be excluded. However, \citet{Ng10} did not include the simultaneous \textit{RXTE} data in their fits, which can affect the Fe emission line profile that strongly depends on the underlying continuum. Triggered by the debate about the Fe line in 4U 1636--53, we obtained and analysed new \textit{XMM-Newton/RXTE} observations of 4U 1636--53, and re-analysed the previous \textit{XMM-Newton/RXTE} data of 4U 1636--53. We investigated and corrected for instrumental effects, used a wide bandpass spectrum (0.8--120 keV), and fit the Fe emission line with the most commonly used phenomenological model and with a relativistically blurred ionised reflection model which self-consistently includes Compton broadening. Additionally, for the first time in Fe line studies, we investigated the contribution of different sources of disc illumination that could produce the reflection spectrum, and their contribution in the different spectral states. 

All together, 4U 1636--53 is an interesting source exhibiting several features that could be used to constrain key parameters like the NS mass and radius, as well as to test the standard NS accretion picture described by the colour-colour diagram. Comparing the properties of the Fe line as a function of spectral state -- or even more specifically with the general properties of the spectrum like luminosity, blackbody temperature of disc and boundary layer -- may help us to constrain the origin of the line emission region. Furthermore, kHz QPOs are seen to vary in frequency depending on the spectral state of the source \citep[see e.g.][]{Belloni07,Altamirano08,Sanna12}. Therefore, connecting the position in the CD with the Fe line properties and the kHz-QPO frequencies, may help to break the degeneracy in the models used to explain the spectral and timing features in accreting NS systems. In this paper, however, we focussed on the spectral analysis and the Fe line properties as this requires a careful investigation. In a companion paper (Sanna et al. 2013, in prep.) we combine the results of the spectral analysis presented in this paper with the timing analysis of the simultaneous \textit{RXTE} observations. 
%
%
%
\section{Observations and data reduction}\label{sec:obs}
4U 1636--53 has been observed with \textit{XMM-Newton} nine times in the last decade (between 2000 and 2009). The first two observations were taken with all CCD cameras operated in imaging mode and were strongly affected by pileup. In the other seven observations, of all the CCD cameras only the EPIC-pn \citep[hereafter PN;][]{Strueder01} camera was on, operating in timing mode. In this mode one of the dimensions of the CCD is compressed into one single row to increase the read out speed. In our spectral analysis we only used the \textit{XMM-Newton} data taken with PN in timing mode. We did not include the high-resolution data from the Reflection Grating Spectrometer in the energy range 0.33--2.5 keV. We note that the \textit{XMM-Newton} observation taken on March 14, 2009 had a flaring high-energy background during the full $\sim$40 ks exposure, and we therefore did not include this observation in our spectral analysis. 

Since 1996 \textit{RXTE} observed 4U 1636--53 more than a 1000 times; from March 2005 the source was regularly observed for $\sim$2 ks every two days \citep[for the first results of this monitoring campaign see][]{Belloni07}. For our spectral analysis we only included the ten \textit{RXTE} observations that were taken simultaneously with \textit{XMM-Newton}. In Table~\ref{tab:obs} we give an overview of the observations, labelled in a chronological order. Using the pointed \textit{RXTE}/PCA observations of the last $\sim$5 years, in Figure~\ref{fig:lc} we show the long-term light curve of 4U 1636--53 on which we marked the moments of the \textit{XMM-Newton} observations used in this paper.
\setcounter{table}{0}
\begin{table*}
\begin{minipage}{160mm}
\begin{center}{
\scriptsize \caption{\textit{XMM-Newton}/\textit{RXTE} observations of 4U 1636--53 used in our analysis}\label{tab:obs}
\begin{tabular}{cccccc}

\multicolumn{1}{c}{Observation} & 
\multicolumn{1}{c}{Instrument} & 
\multicolumn{1}{c}{Identification Nr.} &  
\multicolumn{1}{c}{Start Date} &  
\multicolumn{1}{c}{Start Time} &  
\multicolumn{1}{c}{Exposure(ks)$^{\star}$} \\\hline\hline
Obs. 1 & \textit{XMM-Newton} & 0303250201      & 29-08-2005 & 18:24:23 & 25.7 \\
      & \textit{RXTE}       & 91027-01-01-000 &            & 16:35:28 & 26.2 (PCA) \\
      &                     &                 &            &          & 9.0 (HEXTE) \\

Obs. 2 & \textit{XMM-Newton} & 0500350301      & 28-09-2007 & 15:44:56 & 14.3 \\
      & \textit{RXTE}       & 93091-01-01-000 &            & 14:47:28 & 26.9 (PCA) \\
      &                     &                 &            &          & 8.8 (HEXTE) \\

Obs. 3 & \textit{XMM-Newton} & 0500350401      & 27-02-2008 & 04:15:37 & 34.7 \\
      & \textit{RXTE}       & 93091-01-02-000 &            & 03:46:56 & 25.3 (PCA) \\
      &                     &                 &            &          & 8.3 (HEXTE) \\

Obs. 4 & \textit{XMM-Newton} & 0606070201      & 25-03-2009 & 22:59:30 & 23.8 \\
      & \textit{RXTE}       & 94310-01-02-03  &            & 23:00:32 & 1.9 (PCA) \\
      & \textit{RXTE}       & 94310-01-02-04  & 26-03-2009 & 00:39:28 & 1.6 (PCA) \\
      & \textit{RXTE}       & 94310-01-02-05  &            & 02:17:36 & 1.4 (PCA) \\
      & \textit{RXTE}       & 94310-01-02-02  &            & 03:54:24 & 1.3 (PCA) \\
      &                     &                 &            &          & 2.2 (HEXTE)$^{a}$\\
Obs. 5 & \textit{XMM-Newton} & 0606070301      & 05-09-2009 & 01:57:03 & 32.8 \\
      & \textit{RXTE}       & 94310-01-03-000 &            & 01:17:36 & 16.6 (PCA) \\
      & \textit{RXTE}       & 94310-01-03-00  &            & 08:20:32 & 7.3 (PCA)  \\
      &                     &                 &            &          & 7.6 (HEXTE)$^{a}$\\
Obs. 6 & \textit{XMM-Newton} & 0606070401      & 11-09-2009 & 08:48:11 & 21.1 \\
      & \textit{RXTE}       & 94310-01-04-00  &            & 08:42:24 & 18.4 (PCA) \\
      &                     &                 &            &          & 5.7 (HEXTE)$^{a}$\\\hline
\multicolumn{6}{l}{$^{\star}$ Final exposure time after excluding X-ray bursts, detector drops, and background flares; see}\\
\multicolumn{6}{l}{Section~\ref{subsec:XMM} and~\ref{subsec:RXTE} for more details.}\\
\multicolumn{6}{l}{$^{a}$ Total exposure time of the combined HEXTE data; see Section~\ref{subsec:RXTE} for more information.}
\end{tabular}
\normalsize}
\end{center}
\end{minipage}
\end{table*}
\setcounter{figure}{0}
\begin{figure*}
\begin{center}
\resizebox{2\columnwidth}{!}{\rotatebox{270}{\includegraphics[clip]{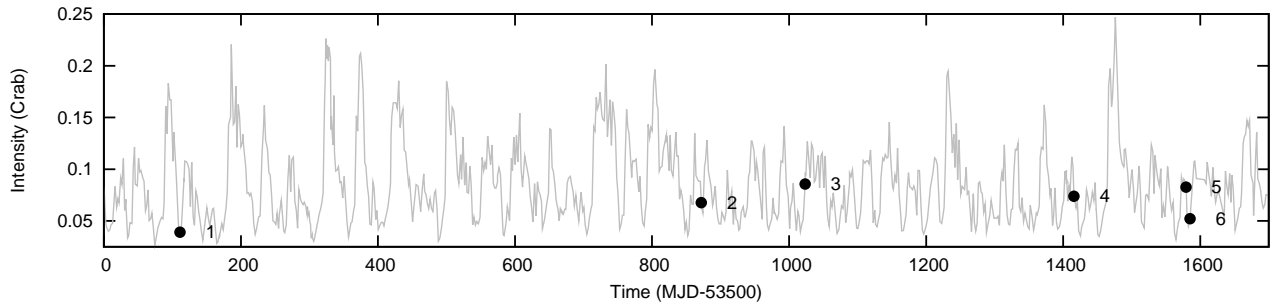}}}
\caption{Long-term light curve of 4U 1636--53 based on pointed \textit{RXTE}/PCA observations of the last $\sim$5 years. The intensity is the 2--16 keV Crab normalised PCA count rate (see Section~\ref{subsec:ccd}). The numbered points mark the six \textit{XMM-Newton} observations, with the numbers corresponding to the observations given in Table~\ref{tab:obs}.}
\label{fig:lc}
\end{center}
\end{figure*}
%
%
%
\subsection{\textit{XMM-Newton} data reduction}\label{subsec:XMM}
For the \textit{XMM-Newton} data reduction we used the tasks of the Science Analysis System (SAS) version 12.1, applying the latest calibration files (xmmsas\_20120523). To generate the concatenated and calibrated PN event files we used the meta-task \texttt{epproc}. We corrected for rate-dependent charge transfer inefficiency (CTI) effects using \texttt{epfast}, and we filtered out X-ray bursts, telemetry drops, and intervals of flaring high-energy background. For the latter, we excluded time-intervals in which the 10--12 keV count rate for single events (\textsc{pattern}=0) was larger than 1~cts~s$^{-1}$. We generated a time-averaged source spectrum per observation, using single and double events only (\textsc{pattern} $\leq 4$), excluding events at the edge of the CCD and at the edge of a bad pixel (\textsc{flag}=0), and using events within a 17-column wide box, centred on the source position, for which the central 3 columns were excised in order to correct for pileup effects (removing $\sim$70\% of the detected photons). We excised the central 5 columns for Obs. 2 and Obs. 3 (removing $\sim$90\% of the detected photons; see Appendix~\ref{app:pile-up} for details). We generated the response matrices using the task \texttt{rmfgen}. Following the recommendations given in the SAS User guide (Chapter 4.6.1)\footnote{\texttt{http://xmm.esac.esa.int}} for piled-up observations in timing mode, we produced the ancillary response files using the task \texttt{arfgen}. For PN timing mode observations, the point spread function (PSF) of the telescope extends further than the CCD boundaries, and extracting the background spectrum from the outer regions of the CCD leads to a spectrum contaminated with source photons \citep[see][for a discussion on contaminated background spectra]{Hiemstra11,Ng10}. To model the background, we extracted a spectrum from the outer region of the CCD (RAWX in 5--20) of a ``blank field'' observation instead \cite[see][]{{Ng10,Hiemstra11}}. Based on similar sky coordinates and column density along the line of sight, we used the timing mode observation of the BH GX 339--4 (Obs 0085680601), in which the source was not significantly detected, as our ``blank field'' for all six \textit{XMM-Newton} observations. Finally, we rebinned the source spectra such that we oversampled the PN energy resolution by a factor of 3 ensuring that we have a minimum of 25 counts per bin.
%
%
%

%
%

%
%
%
\subsection{\textit{RXTE} data reduction}\label{subsec:RXTE}
\subsubsection{Energy spectra}
For the \textit{RXTE} data reduction we used the \textsc{heasoft} tools version 6.10, following the recipes in the \textit{RXTE} cook book\footnote{\texttt{http://heasarc.gsfc.nasa.gov/docs/xte/recipes/cook\_book.html}}. Applying the standard screening criteria and excluding time-intervals of detector drop-outs and X-ray bursts, we used the tool \texttt{saextrct} to extract the PCA spectra from \textrm{Standard-2} data, where we only included events from the third proportional counter unit (PCU2) being this the best-calibrated detector. We corrected for PCA deadtime, and applied a 0.6\% systematic error to the PCA data. The PCA background was estimated with the tool \texttt{pcabackest} and the response files were generated using \texttt{pcarsp}. After excluding detector drop-outs and X-ray bursts, we produced the HEXTE spectra for cluster-B events only, using the script \texttt{hxtlcurv}. With the tool \texttt{hxtrsp} we generated the HEXTE response files. No systematic errors were added to the HEXTE data. For the observations starting on March 25 and September 5, 2009 (Obs. 4 and Obs. 5, respectively), \textit{RXTE} did not cover the full \textit{XMM-Newton} observation but several shorter exposures were taken instead. To speed up the fitting, we used the PCA spectrum with longest exposure, since all the spectra were consistent with each other within errors. However, in order to improve the poor HEXTE statistics we combined the individual HEXTE spectra with the tool \texttt{sumpha}. Net exposures of the final used PCA and HEXTE data are given in the last column in Table~\ref{tab:obs}.
%
%
%
\subsubsection{Intensity and Colours of 4U 1636--53}\label{subsec:ccd}
\begin{figure*}
\begin{center}
\resizebox{2\columnwidth}{!}{\rotatebox{270}{\includegraphics[clip]{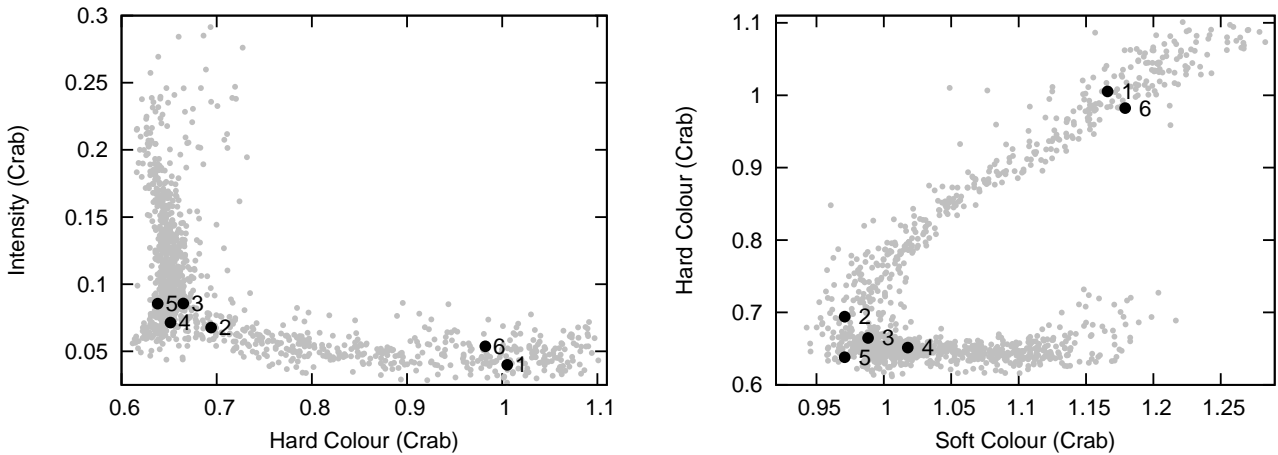}}}
\caption{Hardness-Intensity diagram (\textit{left}) and colour-colour diagram (\textit{right}) of 4U 1636--53. Each grey point represents the averaged Crab-normalised colours and intensity per observation, for a total of 1281 observations (see text in Section~\ref{subsec:ccd} for details). The numbered black points mark the position of the six \textit{XMM-Newton} observations, with the numbers corresponding to the observations given in Table~\ref{tab:obs}.}
\label{fig:ccd}
\end{center}
\end{figure*}
We calculated X-ray colours and intensity of 4U 1636--53 using the \textrm{Standard-2} PCA data. The intensity represents the 2--16 keV count rate, and soft and hard colours are defined as the count rate ratio in the energy bands 3.5--6.0 keV / 2.0--3.5 keV and 9.7--16.0 keV / 6.0--9.7 keV, respectively. To obtain the exact count rate in each of these energy bands we interpolated linearly in channel space. To correct for the gain changes and differences in the effective area between the PCUs, as well as to correct for the differences due to changes in the channel-to-energy conversion of the PCUs as a function of time, we normalised by the Crab Nebula values obtained close in time to our observations \citep[for details see][]{Kuulkers94,Altamirano08}. We finally averaged the normalised colours and intensity per PCU for the full observation using all available PCUs. 

In Fig~\ref{fig:ccd} we show the HID and CD of 4U 1636--53, with each point representing the average intensity, hard and soft colours per observation for a total of 1281 observations, up to May 2010. In these diagrams we also marked the position of the source at the time of the \textit{XMM-Newton} observations. Based on the position of the source in both diagrams, we concluded that Obs. 1 and Obs. 6 were in the hard state, also referred to as the transitional state \citep{Lin07}, characterised by a relatively high hard colour and low X-ray intensity. The positions of Obs. 2--5 were consistent with the soft state \citep[][]{Klis06,Lin07}, where both colours almost reach their minimum values and the X-ray intensity is moderate. Following \citet{Lin07}, in the rest of this paper we refer to Obs. 1 and Obs. 6 as the transitional-state observations, and we refer to Obs. 2--5 as the soft-state observations.
%
%
%
\section{Spectral analysis and results}\label{sec:analysis}
For the spectral fitting of the \textit{XMM-Newton/RXTE} data of 4U 1636--53, we used \textsc{xspec} version 12.7.1 \citep{Arnaud96}. To study the evolution of the Fe emission line and its underlying continuum as a function of the spectral state, we fit each of the individual observations with a set of phenomenological models and a relativistically smeared reflection model in addition to the direct emission components (see following Sections for the model description). In order to study the Fe line properly, we need to have a well-defined continuum on both sides of the Fe bandpass. For that reason we fit the \textit{XMM-Newton/RXTE} spectra in the 0.8--120 keV bandpass, with PN covering the 0.8--11 keV range (from channel 27 up to channel 267, unless otherwise mentioned), PCA taken in the 10--25 keV band (from channel 22 and 53), and HEXTE covering the 20--120 keV range (between channels 11 and 45). Spectral uncertainties are given at 90\% confidence ($\Delta \chi^2=2.706$ for a single parameter).

In all our fits we included a \textsc{phabs} component to account for interstellar absorption, using the abundances and photo-electric cross section of \citet{Wilms00} and \citet{Verner96}, respectively. Additionally, to account for flux calibration disparities between the different instruments, we added a multiplicative constant to the model.
%
%
%
\subsection{Spectral model}\label{sec:the-model}
\subsubsection{Direct emission}\label{subsec:direct}

Spectral modelling of accreting NSs has been, and still is, a controversial matter. A variety of models have been proposed in the past years. All models, however, include at least a soft/thermal and a hard/Comptonized component that vary according with the source state. The soft/thermal emission comes from the accretion disk and from the NS surface, and are typically modelled with a multicolour disk blackbody \citep[][]{Mitsuda84} and a blackbody, respectively. The hard/Comptonized emission is probably created by the up-scattering of soft photons coming either from the accretion disk, the NS surface, or both simultaneously, in the hot electron gas surrounding the central region of the system (usually referred as \textit{corona}). The hard emission is modelled with a power law with or without cut off at high energies, although thermal comptonisation models are also widely used.

In this work we used a multicolour disc blackbody (\textsc{diskbb}) to fit the thermal emission from the accretion disc, which was required by the fits, both in the soft and the transitional state. To model the thermal emission from the NS surface we used a single-temperature blackbody \citep[see][]{Lin07}. For the hard emission we used the thermally comptonized continuum model \textsc{nthcomp} \citep[][]{Zycki89,Zdziarski96}, which compared to an exponentially cutoff power law describes more accurately the high energy shape and the low energy rollover, with similar number of parameters. This model allows to select the spectral shape of the source of seed photons between a (quasi)blackbody and a disk blackbody. During the fitting process we used either of the thermal components as source of seed photons for \textsc{nthcomp}. Both options gave statistically acceptable fits, however when we linked the emission from the NS surface/boundary layer to \textsc{nthcomp} the blackbody emission itself turned out to be negligible. We therefore opted for the disk blackbody as the source of seed photons for the \textsc{nthcomp} component.

Additionally, the emission coming from the inner parts of the accretion disc, close to the NS, may be modified by relativistic effects. For a rotating NS, these effects are similar as for a spinning BH \citep{Miller98} and are approximately described by a Kerr metric characterised by the dimensionless angular momentum, $a_{*}=cJ/(GM^{2})$. For the case of a NS, the metric also depends on the internal structure of the NS, set by the equation of state (EoS), which is unknown. However, the spin parameter for a NS can be approximated as $a_{*}=0.47/P$[ms] \citep{Braje00}. For 4U 1636--53, with $\nu=581$ Hz \citep{Strohmayer02}, the spin parameter is $\sim$0.27. This value is consistent with what \citet{Miller98} found where, for a 500 Hz rotating NS, a given EoS, and a NS mass of 1.4 M$_{\odot}$, the spin parameter is $\sim$0.23. Scaling this value up to a rotation of 581 Hz gives $a_{*} \simeq 0.27$. To test whether the direct disc emission in the spectrum of 4U 1636--53 is significantly affected by relativistic effects we tried, instead of the \textsc{diskbb}, a multi-temperature blackbody model for a thin, general relativistic accretion disc in a Kerr metric, which also includes self-irradiation \citep[\textsc{kerrbb};][]{Li05}. However, using the \textsc{kerrbb}, assuming a distance of 6.5 kpc, a NS mass of 1.4 M$_{\odot}$, and $a_{*}=0.27$, did not significantly improve the fit. Moreover, the uncertainty in the distance and the NS mass of 4U 1636--53, makes the outcome of this model less reliable. Therefore, for the rest of our analysis we only used the \textsc{diskbb} component to fit the direct disc emission.

To summarise: we fitted both the soft and the transitional state of the source with a multicolour disk blackbody (\textsc{diskbb}), plus a single-temperature blackbody (\textsc{bbody}) and the thermally comptonized component (\textsc{nthcomp}) with the soft seed photons coming from the accretion disk.
%
%
%
\subsubsection{Phenomenological and reflection models for the iron emission line}
\subsubsection*{Phenomenological models:}
After fitting the continuum we found several residuals in the whole PN range, with the most prominent one in the energy range 4--9 keV around the Fe line emission region. In order to fit these residuals we added a Gaussian emission line with the energy constrained between 6.4 keV and 6.97 keV (Fe \textit{K} band), while the other parameters of the line (normalisation and width) were free to vary. In all  six observations the data were well fitted by this component, however all the lines were very broad, showing widths ranging from $\sim$1.0 to $\sim$1.4 keV (see Table~\ref{tab:simple_line} for more details on the fit parameters). Triggered by this and by previous results on the Fe emission line in 4U 1636--53  \citep[][]{Pandel08, Cackett10, Ng10}, we used a set of phenomenological models describing relativistically-broadened lines. We selected three of the most commonly used models by the community: \textsc{diskline},  \textsc{laor}, and \textsc{kyrline}.

\textsc{diskline} is a relativistic model for a Schwarzschild metric, $a_{*}=0$, \citep{Fabian89}, thought to be suitable for NS with dimensionless angular parameter lower than 0.3 for which the metric should only marginally deviate from Schwarzschild \citep[see][for details on the subject]{Miller98}. Notice that the \textsc{diskline} model does not include light bending corrections. 

\textsc{laor} is a relativistic line model which includes the effects of a maximally rotating Kerr black hole, $a_{*}=0.998$, \citep{Laor91}, including light bending corrections. Although this model was meant for black holes, it has been largely used to fit disc lines in NS systems \citep[see][for comparison between \textsc{diskline} and \textsc{laor} in the NS LMXB Serpens X-1]{Bhattacharyya07}.

The \textsc{kyrline} component includes all relativistic effects for the Kerr metric \citep[][]{Dovciak04}. Differently from the previous two models, the metric changes with the spin parameter, and can be adjusted for any value of $a_{*}$ between 0 and 0.998. The line profile is calculated for emission from outside the ISCO only, and includes the effects of limb darkening. \\
\subsubsection*{Reflection model:}
When X-rays irradiate an optically thick material such as the accretion disc, they produce a reflection spectrum including fluorescence lines, recombination and other emission \citep[see e.g.][]{Fabian89,Ross93}. When the illuminating flux is high enough (possibly combined with the thermal blackbody radiation intrinsic to the accretion flow) the surface of the accretion disc is expected to be ionised. The ionisation state of the reflecting material determines the shape of the reflection spectrum \citep[e.g.][]{Ross93}, and thus it is important to solve for the thermal balance of the disc. 
In most X-ray sources the incident emission for the reflection spectrum is generally a hard power-law spectrum. However, in NS systems, the emission coming from the NS surface/boundary layer may contribute as well. As the shape of the reflection continuum also depends on the incident emission, it is important to investigate the different possible sources of irradiation.
Several ionised reflection models are publicly available within \textsc{xspec}, which self-consistently compute the ionisation and thermal balance according to the radiation field. In our analysis we used simultaneously two different reflection models:  \textsc{rfxconv} and \textsc{bbrefl}.

\textsc{rfxconv} is an updated version of the code in \citet{Done06}, using the \citet{Ross05} reflection \texttt{atables}. This model can be used with any input continuum and has therefore the advantage of not having a fixed exponential cut-off in the illuminating power law at 300 keV as in the model of \citet{Ross05}. We decided not to reflect the blackbody emission with \textsc{rfxconv} because this component calculates the ionisation balance assuming that the ionising spectrum is  a power law, and this is not appropriate for the case of the reflection of blackbody emission off the disc. Instead, we used \textsc{bbrefl}, which provides the reflection spectrum from a constant density disc illuminated by a blackbody \citep[][]{Ballantyne04}. Both components also include lines and edges of the most important elements and ions which are Compton and thermal broadened depending on the ionisation state of the reflecting material.

Since the reflection spectrum may be further smeared by relativistic effects in the inner regions of the accretion disc, we convolved the reflected emission with a Kerr-metric kernel which includes relativistic effects \citep[\textsc{kerrconv};][]{Brenneman06}, where we fixed the spin parameter to 0.27.
In this work we explored how the corona and the boundary layer contributed individually to the reflection spectrum. Details on the specific parameters and their settings are given in Section~\ref{sec:parameterSetting}, and the results are discussed in Section~\ref{sec:results}.

%
%
%
\subsubsection{Residual absorption and emission features}\label{subsec:residuals}
After fitting the spectra with the continuum model plus the relativistic Fe line profile, the \textit{XMM-Newton} data still showed some narrow residuals. especially at low energies. Besides the commonly seen residuals at the instrumental Si-K and Au-M edges at $\sim$1.8 keV and $\sim$2.2 keV, respectively, there were a few other clear features apparent in the residuals. All six observations showed absorption features at $\sim$0.9 keV, at $\sim$1.5 keV and at $\sim$ 9.2 keV. Taking into account the moderate PN energy resolution ($\sim$70 eV at 1 keV), the $\sim$0.9 keV feature could be due to absorption by Ne IX (0.905 keV) or Fe XVIII (0.873 keV), and may possibly be of astrophysical origin. The energy of the $\sim$1.5 keV feature is either consistent with the Fe XXI--XXIV blend or with the Al-K edge. The effective area curves of the PN show a strong Al-K edge at $\sim$1.56 keV, which suggests that this feature is likely an artefact of the PN calibration. The absorption feature at $\sim$9.2 keV is consistent with the Fe XXVI edge. 
The above residuals contributed to an enhanced $\chi^{2}$. To improve the fit we modelled these residuals with a Gaussian absorption or emission component. Since the main focus of this paper is the spectrum at energies near and above the energy of the Fe line, we do not discuss these lines further. In Obs.~5 we further ignored 11 channels from the PN data (spread between channels 50 and 148, corresponding to the energy range $\sim$ 1.3 to $\sim$ 4.7 keV), that appeared in the spectrum like very narrow absorption or emission features.
%
%
%

\subsubsection{Fitting procedure}

We started fitting the energy range 0.8-100 keV using PN (0.8-10 keV), PCA (3-20 keV) and HEXTE (20-100 keV) data. From the fits with the phenomenological model \textsc{kyrline} for the line (see Sec. 3.1.1 for details about the model) we noticed a mismatch between PN and PCA data between 4 and 8 keV. To investigate the mismatch we allowed the phenomenological model of the line to vary between PN and PCA, and we found that, except for the normalisation (with the iron line being a factor $\sim$ 4 stronger in the PN than in the PCA spectra), the parameters of the Fe line were consistent within errors. By letting the line normalisation free between the two instruments the fit improved significantly. The values of the normalisation in PN and PCA were, in both cases, significantly different from zero. We tested the fits excluding both PCA and HEXTE data, and we fitted the model previously described to the PN data only. We found that, for the observations in the transitional state the line parameters do not change much between PN, and PN+PCA+HEXTE, except for the normalisation of the line and the inner radius of the disc that are marginally larger in the case we used the three instruments. All continuum components were statistically required to fit the PN-only spectrum. In the soft spectra all the line parameters were consistent within the errors with being the same. In this case the direct emission from the corona is not statistically required by the fit. This is however understandable given that the PN only extends up to 10 keV and the hard emission is less important in the soft than in the transitional state.

Based on these comparisons, and to be able to properly constrain the underlying continuum and model the Fe line emission, we proceeded further by including all three instruments in our fits, excluding the PCA data below 10 keV, where we relied on the PN data. We further modelled the residual absorption between 9 and 10 keV using an edge with energy free between 9.2 and 9.3 keV, which could be due to Fe XXVI. The optical depth of the edge varied between 0.05 and 0.1. Not adding the edge to the fits does not affect the parameters of the line. The high level of ionisation required to create this feature contrasts with the Fe line energy values found from the fits, suggesting that this absorption edge might be due to a calibration mismatch between PN and PCA.

\subsection{Parameter settings}\label{sec:parameterSetting}

\subsubsection{Continuum}
As we mentioned,  we used the \textsc{diskbb}, \textsc{bbody} and \textsc{nthcomp} components to model the direct emission coming from the accretion disc, boundary layer, and corona, respectively. The parameters for these components are the disc temperature at the inner edge, $kT_{\rm in}$, the temperature of NS surface, $kT_{\rm BB}$, and the Comptonisation photon index, the electron temperature (high energy rollover), and the seed photon temperature (low energy rollover), $\Gamma$, $kT_{\rm e}$, and $kT_{\rm dbb}$, respectively. $N_{\rm dbb}$, $N_{\rm BB}$, and $N_{\rm NTH}$ are the normalisation parameters for the \textsc{diskbb}, \textsc{bbdoy} and \textsc{nthcomp} components, respectively. All the direct emission parameters were free to vary (except $kT_{\rm dbb}$ which was set equal to $kT_{\rm in}$ of \textsc{diskbb}), but were coupled between the different instruments.

\begin{figure}
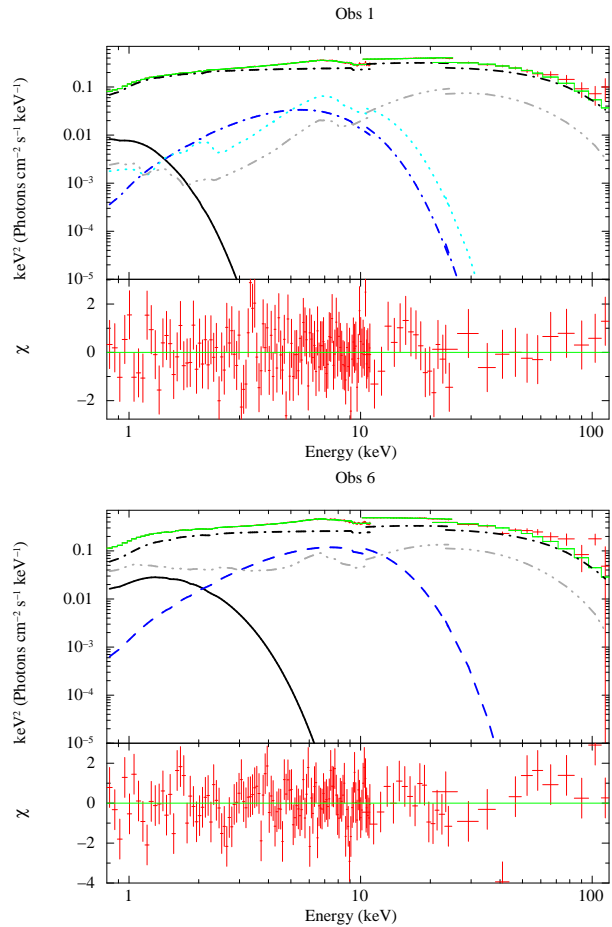

\begin{center}
\resizebox{\columnwidth}{!}{\rotatebox{270}{\includegraphics[clip]{ob1_model_paper_final.ps}}}
\resizebox{\columnwidth}{!}{\rotatebox{270}{\includegraphics[clip]{obs6_model_paper_final.ps}}}
\caption{The two transitional-state spectra (\textit{top}, Obs. 1, and \textit{bottom}, Obs. 6) of 4U 1636--53. Each plot shows the simultaneously fitted \textit{XMM-Newton/RXTE} spectra and unfolded model in the main panel, and the residuals in units of sigmas in the sub-panel. To account for the direct emission from the accretion disc, NS surface/boundary layer, and corona, this model contains a \textsc{diskbb} (black/line), \textsc{bbody} (blue/dashed), and \textsc{nthcomp} (black/dashed-dotted) component, respectively. The relativistically smeared reflected emission is due to two incident components: the NS surface/boundary layer (cyan/dotted) and the corona (light grey/dashed-triple dotted).}
\label{fig:spectra_hard}
\end{center}
\end{figure}


\subsubsection{Phenomenological line models}
The phenomenological line models are parameterised by the inclination angle of the accretion disc, $i$, the rest energy of the line, $E_{\rm line}$, the inner and outer edge of the disc, $R_{\rm in}$ and $R_{\rm out}$, the emissivity index, $\beta$ (\textsc{kyrline} has the option to allow the emissivity index to be different in the inner and outer disc regions; here we chose to have a disc described by a single emissivity index), and the normalisation of the line, $N_{\rm line}$, in photons cm$^{-2}$ s$^{-1}$. We constrained $E_{\rm line}$ to range between 6.4 and 6.97 keV, and fixed the outer disc radius to be 1000 $GM/c^{2}$. Using \textsc{kyrline} we fitted the data fixing  $a_{*}$ to three different values: 0, 0.27 and 0.998, and limited the inner radius to the ISCO $R_{\rm in} \gtrsim$ 6, 5.12, 1.23 $GM/c^{2}$, respectively. 


\subsubsection{Reflection model}
\label{subsec:reflModel}
The ionised reflection emission is characterised by three parameters: the scaling reflection factor, $\Omega_{\rm refl}$,  from \textsc{rfxconv} ($\Omega_{\rm refl}<0$ represents only the reflected component), \textsc{bbrefl} normalisation (flux per emitting area), and the ionisation parameter, $\xi=4\pi F/n_{\rm H}$, with $F$ the total illuminating flux, and $n_{\rm H}$ the hydrogen number density. Under the assumption that the reflection region in the accretion disc has a constant ionisation parameter we coupled the ionisation parameter between \textsc{rfxconv} and \textsc{bbrefl}.\\
The relativistic effects, determined by the metric and the disc properties, were included using the \textsc{kerrconv} component which is parameterised by the spin parameter, $a_{*}$, the disc inclination, $\theta$, the disc inner and outer radius, $R_{\rm in}$ and $R_{\rm out}$ (in units of the marginal stable radius, $R_{\rm ms}$), and the disc emissivity index, $\beta$. The latter could be different for the inner and outer disc, although we chose to have a disc described by a single emissivity index. $R_{\rm in}$, $\theta$, and $\beta$ were free to vary, but we fixed $R_{\rm out}$ to its default value (400 $R_{\rm ms}$) and $a_{*}$ to 0.27 (see Section~\ref{subsec:direct}). Given this spin parameter, using equation (3) in \citet{Miller98}, the ISCO is at $\sim$5.12 gravitational radii, $R_{\rm g}$ ($R_{\rm g}=GM/c^2$, this is the first-order approximation valid for spin frequencies below $\sim$400 Hz). All free parameters in the relativistically smeared reflection model were coupled between the different instruments. 

The inclination in \textsc{rfxconv} is given as a cosine function and we fixed $\cos\theta$ to a value consistent with the inclination angle in \textsc{kerrconv} obtained from the initial fits of the relativistically smeared reflection model. For all observations this initial value for the inclination was in the range 63--69$^{\circ}$, which is quite high but still consistent with the results of \citet{Casares06}. However, for most of the observations (except Obs. 2 and 4) we found that the best-fitting values for the inclination inferred from \textsc{kerrconv} were larger than 80$^{\circ}$, which is too high as no eclipses have been seen in this source. In Section~\ref{sec:caveats} we examine this issue in more detail. 
Besides the parameters we directly got from the blurred reflection models, it is also common to provide the strength of the Fe line, usually expressed in terms of its equivalent width (EW). However, as the Fe line is part of the reflection spectrum, it is not possible to separate the emission line from the underlying reflection continuum. Therefore, we estimated the equivalent width of the reflection continuum in the region of the Fe line. As the line is broadened, we used the 4--9 keV range as our line region for all six observations. The equivalent width is defined as \textsf{flux/contin}; with \textsf{flux} being the (unabsorbed) 4--9 keV flux of the reflection spectrum, and \textsf{contin} the average (unabsorbed) 4--9 keV flux density of the direct continuum. In Table~\ref{tab:reflection} we give the equivalent width obtained in this way. However, we note that these values probably overestimate the true equivalent width of the Fe line since \textsf{flux} not only includes the line photons but also the photons of the underlying reflection continuum.

\subsection{Sources of irradiation}\label{sec:irradiation}
\begin{table*}
\begin{minipage}{180mm}
\begin{center}
\tabcolsep=2.0mm
{
\caption{Best-fitting results for the reflection model.}\label{tab:reflection}
\resizebox{1\linewidth}{!}{
\begin{tabular}{clcccccc}\hline

\multicolumn{1}{c}{Component} & 
\multicolumn{1}{l}{Parameter} & 
\multicolumn{1}{c}{Obs. 1} &  
\multicolumn{1}{c}{Obs. 2} &  
\multicolumn{1}{c}{Obs. 3} &  
\multicolumn{1}{c}{Obs. 4} &  
\multicolumn{1}{c}{Obs. 5} &  
\multicolumn{1}{c}{Obs. 6}\\\hline
\textsc{phabs}   & $N_{\rm H}$ ($10^{22}$)     & 0.40$\pm$0.01           & 0.37$\pm$0.03        & 0.39$\pm$0.02                & 0.31$\pm$0.01                 & 0.28$\pm$0.02               & 0.37$\pm$0.01 \\
\textsc{diskbb}  & $kT_{\rm in}$ (keV)         & 0.19$\pm$0.01           & 0.68$\pm$0.03              & 0.72$\pm$0.01                & 0.70$\pm$0.03                 & 0.80$\pm$0.01                &0.40$\pm$0.05 \\
                 & $N_{\rm dbb}$               & 5026$^{+21876}_{-520}$     & 87$^{+51}_{-26}$                     & 74$^{+3}_{-54}$            & 35$^{+22}_{-2}$               & 123$^{+3}_{-4}$              &259$\pm$78 \\
                 & F$_{\rm d}$ ($10^{-11}$)     & 3.9$^{+17.1}_{-0.9}$   & 32.9$^{+20.0}_{-11.1}$         & 35.1$^{+1.5}_{-25.6}$       & 15.1$^{+9.3}_{-1.4}$      & 37.2$^{+3.1}_{-2.4}$       & 9.5$^{+7.2}_{-5.2}$\\
\textsc{bbody}  & $kT_{\rm BB}$ (keV)          & 1.40$^{+0.02}_{-0.03}$           & 1.95$\pm$0.02       & 1.87$^{+0.04}_{-0.01}$               & 1.74$^{+0.02}_{-0.04}$          & 1.37$^{+0.02}_{-0.23}$                &1.89$\pm$0.02 \\
      & $N_{\rm BB}$ ($10^{-3}$)&  0.9$^{+0.2}_{-0.1}$     & 5.7$\pm$0.4           & 2.9$^{+1.0}_{-0.4}$          & 6.2$^{+0.5}_{-0.2}$           & 3.0$^{+3.1}_{-0.2}$          & 3.1$^{+0.5}_{-0.6}$ \\
                 & F$_{\rm b}$ ($10^{-11}$)    & 7.4$^{+1.5}_{-0.2}$  & 48.1$\pm$0.8      &24.5$^{+8.4}_{-3.4}$       & 52.4$^{+4.1}_{-1.8}$        & 21.3$^{+22.0}_{-1.3}$      & 26.2$^{+4.2}_{-5.1}$\\
\textsc{nthcomp}& $\Gamma$                    & 1.94$\pm$0.13           & 2.42$^{+0.20}_{-0.27}$        & 2.46$\pm0.19$       & 2.90$\pm0.04$         & 2.00$^{+0.76}_{-0.04}$        & 1.95$^{+0.04}_{-0.14}$ \\
                 & $kT_{e}$ (keV)         & 17.9$^{+3.1}_{-6.1}$                & 9.5$^{+0.9}_{-0.8}$           & 4.5$^{+0.3}_{-0.2}$          & 16.5$^{+35.9}_{-1.9}$           & 3.0$^{+0.3}_{-0.2}$                 & 16.4$^{+6.7}_{-4.5}$ \\
                 & $N_{\rm NTH}$            & 0.21$^{+0.01}_{-0.03}$           & 0.12$^{+0.08}_{-0.03}$        & 0.36$^{+0.04}_{-0.08}$       & 0.31$\pm0.02$        & 0.15$^{+0.11}_{-0.01}$              &0.18$\pm$0.05  \\
                 & F$_{\rm NTH}$ ($10^{-9}$)    & 1.8$^{+0.4}_{-0.7}$         & 0.9$^{+0.6}_{-0.2}$  &  2.4$^{+0.3}_{-0.6}$   & 1.8$^{+0.2}_{-0.1}$          & 0.8$^{+0.7}_{-0.1}$        & 1.8$^{+0.5}_{-0.8}$ \\
\textsc{kerrconv}& $\beta$                     & 2.8$^{+0.2}_{-0.1}$             & 3.1$^{+0.5}_{-0.3}$            & 4.1$^{+1.2}_{-0.6}$                  &2.5$\pm$0.1                     & 2.8$^{+0.2}_{-0.1}$                 & 4.5$^{+0.0*}_{-2.5}$ \\
                 & $\theta$ (deg)              & 83.1$^{+0.2}_{-2.7}$         & 48.0$^{+3.5}_{-2.6}$         & 85.5$^{+0.6}_{-1.0}$         & 70.3$^{+0.5}_{-1.9}$        & 87.1$^{+0.1}_{-1.5}$        & 88.7$^{+1.1}_{-0.6}$ \\
                 & $R_{\rm in}$ (GM/c$^2$)  & 12.6$^{+1.5}_{-1.8}$  & 7.8$^{+3.1}_{-2.7*}$        & 15.4$\pm$2.7       & 5.6$^{+2.2}_{-0.3}$        & 5.4$\pm$0.1       &19.1$^{+7.6}_{-10.8}$ \\
\textsc{rfxconv} & $\Omega_{\rm refl}$         & -0.98$^{+0.03}_{-0.12}$  & -2.1$^{+0.2}_{-1.6}$ & -0.25$^{+0.15}_{-0.12}$ & - &-0.92$^{+0.1}_{-1.88}$ &-1.29$^{+0.2}_{-0.4}$ \\
                 & F$_{\rm rfx }$ ($10^{-10}$)    & 2.1$^{+0.8}_{-1.5}$  & 11.9$\pm$9.5      & 1.9$^{+1.4}_{-1.3}$       & -          & 0.8$^{+0.5}_{-1.7}$         & 5.6$^{+2.1}_{-3.5}$\\                 

\textsc{bbrefl} & disknorm ($10^{-24}$)& 1.5$\pm$0.1 & -       & 0.013$\pm$0.005      & 1.72$^{+0.03}_{-0.34}$        & 5.79$^{+0.06}_{-0.08}$               & - \\
                 & $\log \xi$                  & 1.00$^{+0.02}_{-0.00*}$ & 3.75$^{+0.0*}_{-0.5}$        & 2.87$^{+0.16}_{-0.30}$       & 1.21$\pm$0.01        & 1.00$^{+0.03}_{-0.00*}$               & 2.49$\pm$0.04 \\
                 & F$_{\rm bbrefl }$ ($10^{-10}$)    & 1.1$\pm$0.1 &  -  & 3.4$^{+2.4}_{-2.7}$       &  2.6$^{+0.1}_{-0.6}$         & 4.1.$^{+3.9}_{-0.2}$         & -\\\\     
           & $\chi^{2}_{\nu}$ ($\chi^{2}/\nu$) & 1.12 (313/278)          & 1.03 (288/279)                & 1.03 (286/278)               & 1.08 (301/279)                & 1.07 (287/268)              & 1.14 (320/279) \\\\
           & Total flux ($10^{-9}$)        & 1.9$^{+0.5}_{-0.8}$ & 2.9$^{+1.1}_{-0.9}$      & 3.0$^{+0.5}_{-0.7}$      & 2.4$\pm$0.2       & 2.6$^{+0.8}_{-0.2}$       & 2.7$^{+0.5}_{-0.9}$\\\\
           & EW (keV)                &  0.21$^{+0.04}_{-0.06}$                 & 0.63$^{+0.47}_{-0.31}$                       & 0.41$^{+0.30}_{-0.28}$                    & 0.18$^{+0.02}_{-0.04}$                     & 0.36$^{+0.29}_{-0.05}$                     & 0.17$\pm$0.09\\\hline\hline
\end{tabular}}
\normalsize}
\end{center}
NOTES.--~A * denotes that the error has reached the maximum or minimum allowed value of the parameter. Uncertainties are given at a 90\% confidence level. $\Omega_{\rm refl} < 0$ represents reflection only, where the actual reflection normalisation is $|\Omega_{\rm refl}|$. $N_{\rm dbb}$ is defined as $(R_{in}/D_{10})^2\cos \theta$, with $R_{in}$ in km, $D_{10}$ the distance in 10 kpc, and $\theta$ the inclination angle of the disc. $N_{\rm BB}$ is $L_{39}/D_{10}^2$, where $L_{39}$ is the luminosity in units of $10^{39}$ erg s$^{-1}$. $N_{\rm NTH}$ is in units of photons keV$^{-1}$ cm$^{-2}$ s$^{-1}$ at 1 keV. All the flux values reported represent the unabsorbed flux in the energy range 0.5-130 keV, all in erg~cm$^{-2}$~s$^{-1}$. EW represents the equivalent width of the reflection component in the 4-9 keV energy range.
\label{reflection}
\end{minipage}
\end{table*}

\begin{figure*}
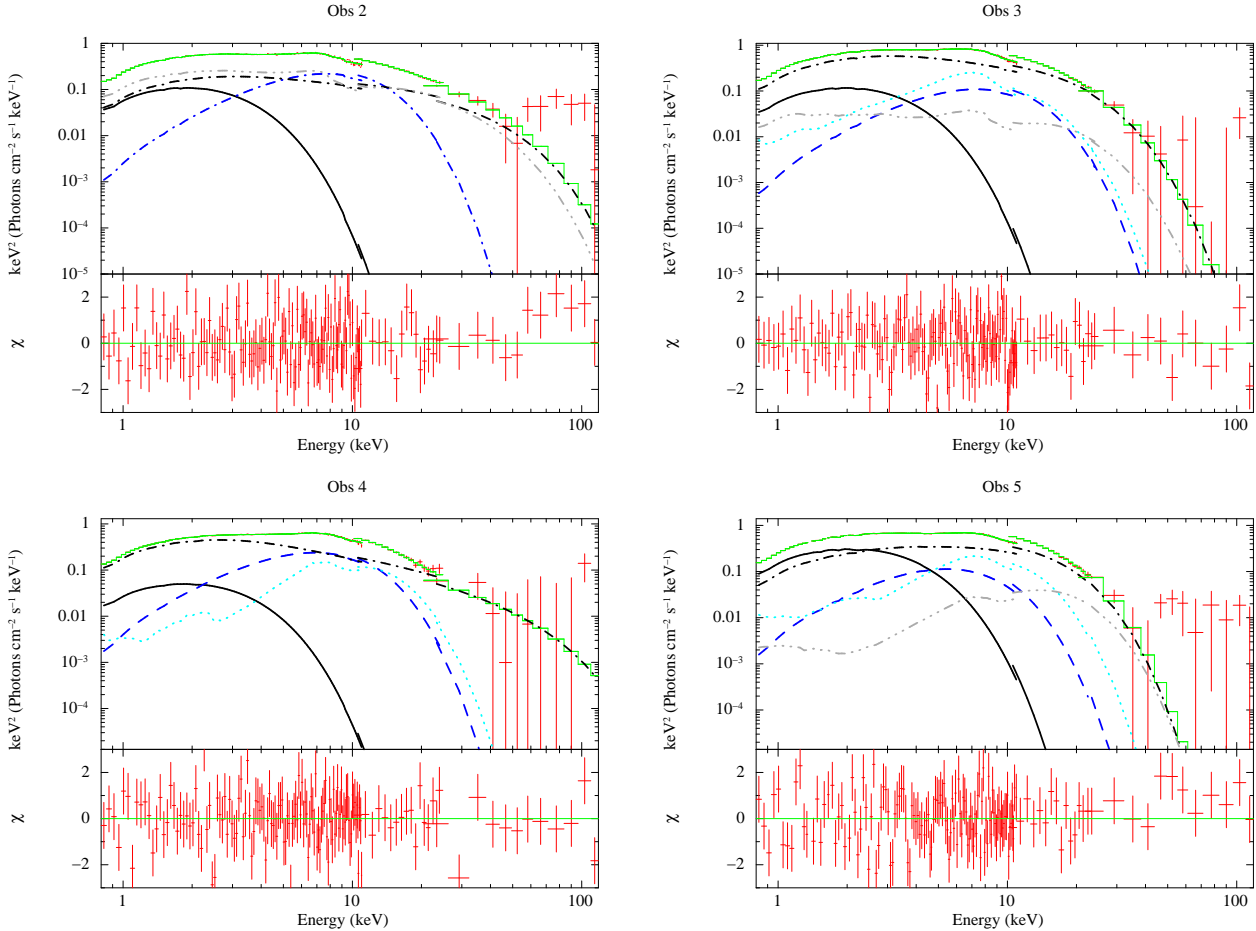

\begin{center}$
\begin{array}{cc}
\includegraphics[scale=0.31,angle=270]{obs2_model_paper_final.ps} & 
\includegraphics[scale=0.31,angle=270]{obs3_model_paper_final.ps} \\
\includegraphics[scale=0.31,angle=270]{obs4_model_paper_final_copy.ps} & 
\includegraphics[scale=0.31,angle=270]{obs5_model_paper_final.ps}\\
\end{array}$
\end{center}
\caption{The four soft-state spectra (\textit{top-left}, Obs. 2, \textit{bottom-left}, Obs. 4, \textit{top-right}, Obs. 3, and \textit{bottom-right}, Obs. 5) of 4U 1636--53. Each plot shows the simultaneously fitted \textit{XMM-Newton/RXTE} spectra and unfolded model in the main panel, and the residuals in sigma units in the sub panel}. Lines and colours are the same as in Figure~\ref{fig:spectra_hard}
\label{fig:refl_soft_spec}
\end{figure*}

To investigate how the boundary layer and the corona contributed to the reflection spectrum in 4U 1636--53, we defined our model as:  {\sc phabs*(diskbb+ bbody + nthcomp + kerrconv*(bbrefl + rfxconv*nthcomp))} (hereafter, model 1), where both the \textsc{bbody} and \textsc{nthcomp} components were reflected and relativistically smeared. We found acceptable fits for the observations, with the reduced $\chi^{2}$ ranging from 1.06 to 1.12 for 278 d.o.f. However, for some observations, we found that one of the two reflection components was not contributing significantly to the total spectrum. We therefore investigated the scenario where only one of the two incident components, \textsc{bbody} or \textsc{nthcomp} (hereafter model 2 and 3, respectively), contributed to the reflected emission. 

In Table~\ref{tab:modelsRefl} we summarised the values of $\chi^{2}$ and number of degrees of freedom for the different fitting models applied. For three out of the 6 observations (Obs.~1, 3 and 5) the best fit was obtained by including both reflection from the NS surface/boundary layer and the corona (model 1). For Obs. 2 and 4, both model 1 and 2 showed the same $\chi^{2}$ with only 1 d.o.f. of difference. $\Omega_{\rm refl}$, in Obs.~2 and 4, was consistent with being zero, hence \textsc{rfxconv$\ast$nthcomp} in model 1 was not required to model the data. For Obs.~4 we then decided to use model 2. For Obs.~2, on the other hand, we finally opted for model 3. Fitting results from model 2 showed that the direct emission from the NS surface/boundary layer was consistent with zero within errors. 
This would imply that the NS surface/boundary layer, although not directly visible, still contributes to the reflection spectrum. While this is in principle possible, it is more likely that the model used to fit this spectrum has too many components, resulting in an over-parameterisation of the problem. Taking all this into account, for Obs.~2 we discarded models 1 and 2, although statistically preferred, and we selected model 3 instead. The case of Obs.~6, was very similar to that of Obs.~2 (see \ref{tab:modelsRefl}). Also in this case, fitting model 1 we found that only the \textsc{bbrefl} component was significantly contributing to the reflected emission, while switching to model 2 the direct emission from the NS surface/ boundary layer was not significantly required. As for Obs.~2, we decided to use model 3. In Figure \ref{fig:spectra_hard} we show the spectra and the individual components of the best-fit models for the transitional observations fitted with the reflection model. The top and bottom panels of Figure \ref{fig:spectra_hard} show Obs. 1 and 6, respectively. As previously mentioned, in Obs. 1 both the reflection from the NS surface/boundary layer (cyan-dotted line) and corona (light grey dashed-triple dotted line) contribute significantly, with the latter accounting for $\sim$30\% of the reflection emission in the 4.0--9.0 keV range. On the other hand, in Obs. 6 the NS surface contribution to the reflection spectrum is not required. \\
\noindent
In Figure \ref{fig:refl_soft_spec} we show spectra and individual components of the best-fit models for the soft-state observations (\textit{top-left}, Obs.~2, \textit{top-right}, Obs.~3, \textit{bottom-left}, Obs.~4, and \textit{bottom-right}, Obs.~5).
Obs.~2 is the only soft-state observation where the reflected spectrum is entirely described by reflection from the corona (light grey dashed-triple dotted line).
Obs.~3, as well as Obs.~5, show contribution from both reflection components, with the the corona (light grey dashed-triple dotted line) contributing $\sim$15\% and $\sim$13\%, respectively in the Fe line region. Finally, Obs.~4 is the only soft-state observation where the corona does not contribute to the reflected spectrum.
%
%
\subsection{Evolution of continuum parameters and line}\label{sec:results}

\begin{figure}
\begin{center}
\includegraphics[scale=0.43]{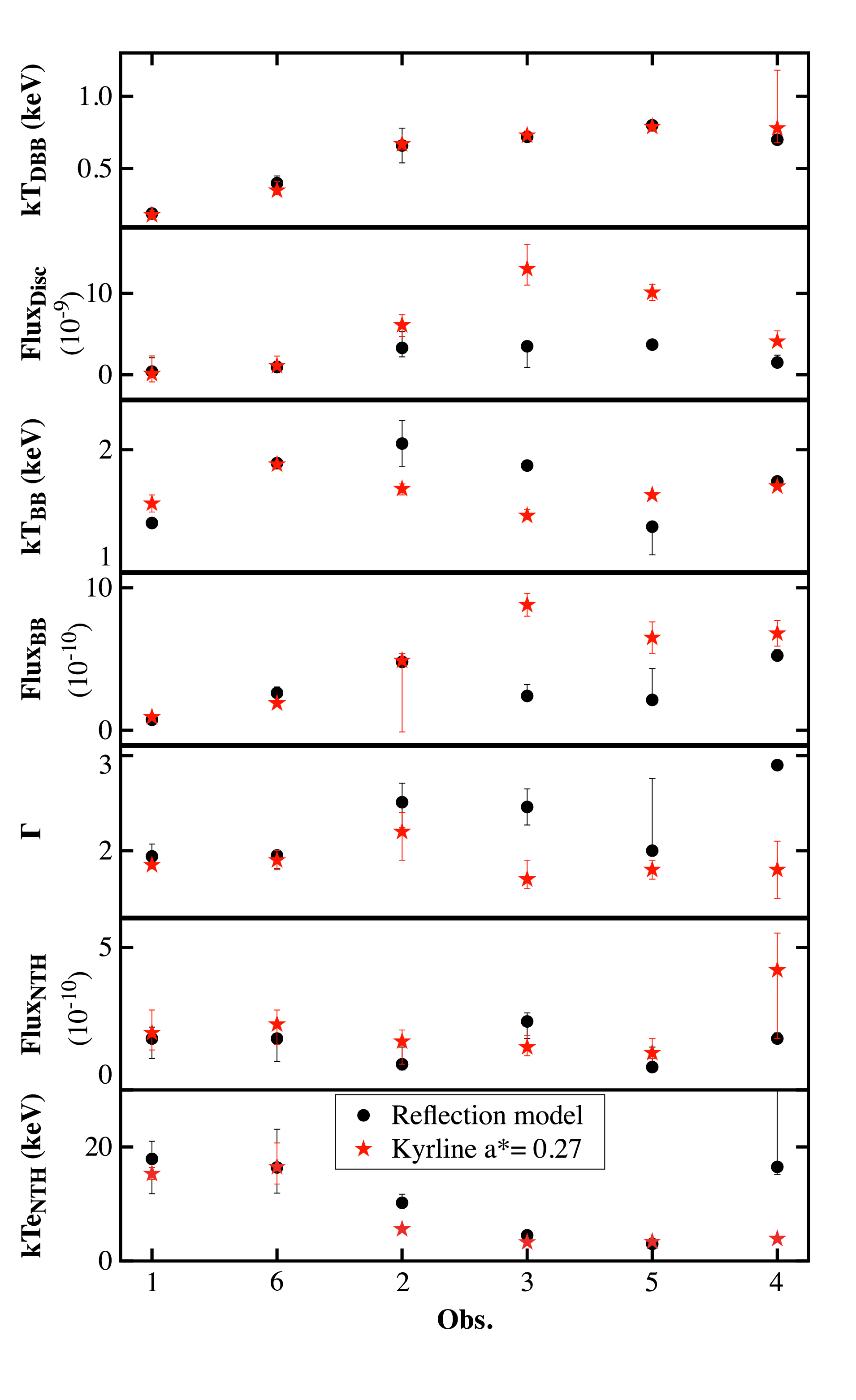}
\caption{Evolution of several spectral parameters as the source moves through the CD. The observations are ordered according to their $S_{\rm z}$ coordinate. From top to bottom, the panels show respectively the evolution of the \textsc{diskbb} temperature, the disc flux, the \textsc{bbody} temperature, the blackbody flux,  \textsc{nthcomp} photon index, \textsc{nthcomp} flux and \textsc{nthcomp} electron temperature. The fluxes are unabsorbed in the 0.5--130 keV range in erg~cm$^{-2}$~s$^{-1}$.}
\label{fig:continuum_prop}
\end{center}
\end{figure}

\subsubsection{Continuum parameters}
The continuum parameters from the reflection and phenomenological models are shown in Table~\ref{tab:reflection} and Tables~\ref{tab:simple_cont1}-\ref{tab:simple_cont2}, respectively.
In order to study the evolution of the individual spectral components in relation to the position of the source in the CD, we ordered the observations according to their colour-colour coordinate $S_{\rm z}$, which is thought to be a function of the mass accretion rate \citep[see e.g.,][]{Kuulkers94}. The order of the six \textit{XMM-Newton/RXTE} observations according with their $S_{\rm z}$ value is: 1--6--2--3--5--4. Applying this order, in Figure~\ref{fig:continuum_prop} we show how several of the spectral parameters changed with the position in the CD, considering both the reflection model and a simple relativistic emission model (\textsc{kyrline} with angular parameter fixed to 0.27). From the first panel of Figure~\ref{fig:continuum_prop} we found that the disc temperature increased as the source became brighter, consistent with the standard accretion disc model \citep[see e.g.,][and reference therein]{Done07}. The disc flux (second panel) did not show a similar trend as the disc temperature, however it is apparent that the flux is on average lower in the transitional-state (Obs.~1 and 6) than in the soft state; although that is more evident when using the phenomenological model (red stars), it can still be noticed for the reflection model. From panels 3 and 4, we found that the flux and temperature of the NS surface/boundary layer did not show any clear trend with $S_{\rm z}$ in the fits with the reflection model. Notice, however, that the blackbody flux in the phenomenological model showed a clear increase between the two source states. The \textsc{bbody} behaviour across the CD does not seem to follow the standard scenario, for which an increase in mass accretion rate should heat up the surface of the NS. Panels 5 to 7 show the corona emission properties across the CD. Both, photon index and flux of the component did not vary systematically as the source moved from transitional to soft state. Notice, however, $\Gamma$ from the reflection model in average larger in the soft state. On the other hand, the electron temperature ($kT_e$) of the scattering cloud clearly decreased as mass accretion rate increased (with the exception of Obs.4 in the reflection model). The latter result is consistent with the scenario in which, as the source moves from transitional to soft state, the accretion disc becomes hotter emitting a larger number of soft photons that will cool down the electrons in the corona.

\subsubsection{Emission line: phenomenological models}

\begin{figure}
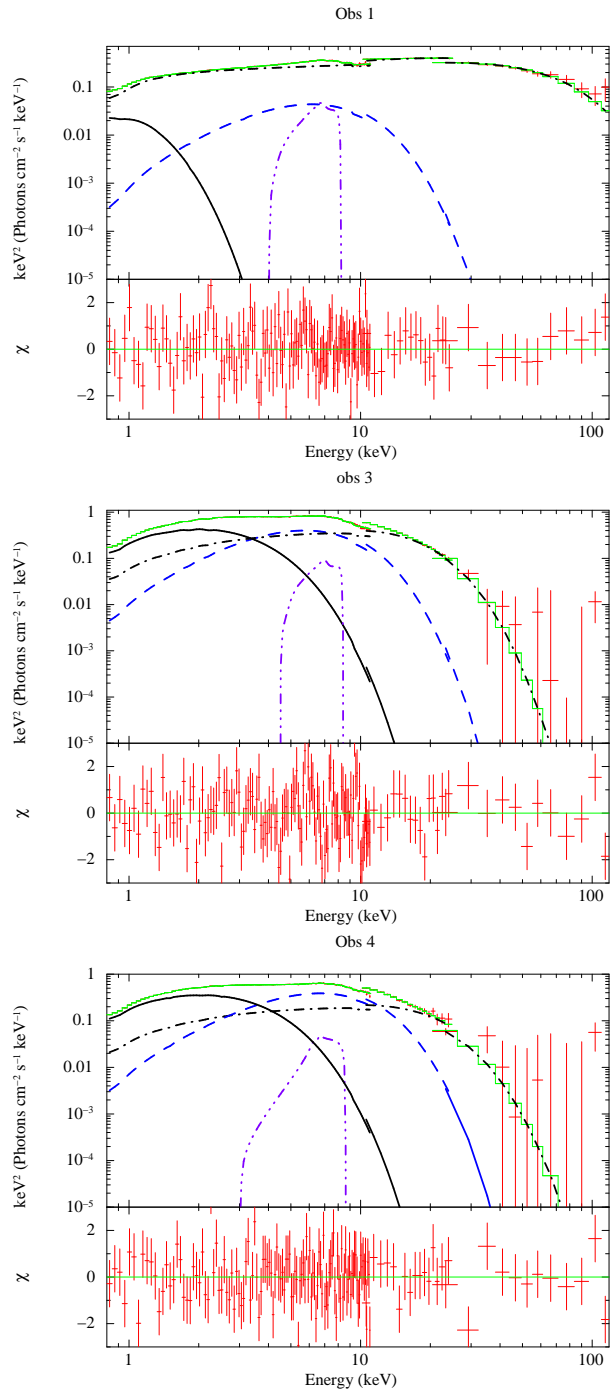

\begin{center}
\resizebox{\columnwidth}{!}{\rotatebox{270}{\includegraphics[clip]{obs1_kyrline_paper2.ps}}}
\resizebox{\columnwidth}{!}{\rotatebox{270}{\includegraphics[clip]{obs3_kyrline_paper2.ps}}}
\resizebox{\columnwidth}{!}{\rotatebox{270}{\includegraphics[clip]{obs4_kyrline_paper.ps}}}
\caption{Transitional-state spectrum (\textit{top}, Obs. 1) and two soft-state spectra (\textit{middle}, Obs. 3  and \textit{bottom}, Obs. 4) of 4U 1636--53. Each plot shows the simultaneously fitted \textit{XMM-Newton/RXTE} spectra and unfolded model in the main panel, and the residuals in sigma units in the sub panel}. The continuum emission components are represented as in Figure~\ref{fig:spectra_hard}. The relativistically smeared Fe emission line is fitted with the phenomenological model \textsc{kyrline} with spin parameter $a_*=0.27$ (violet/dashed-triple dotted).
\label{fig:phen_spec}
\end{center}
\end{figure}

The line parameters from the phenomenological models are given in Table~\ref{tab:simple_line}. In Figure~\ref{fig:phen_spec} we show three representative spectra and individual components of a transitional-state observation (\textit{top}, Obs.~1), and two soft-state observations (\textit{middle}, Obs.~3, and \textit{bottom}, Obs.~4) fitted with relativistic model \textsc{kyrline} with $a_*=0.27$. 

\begin{figure*}
\begin{center}
\includegraphics[scale=0.47]{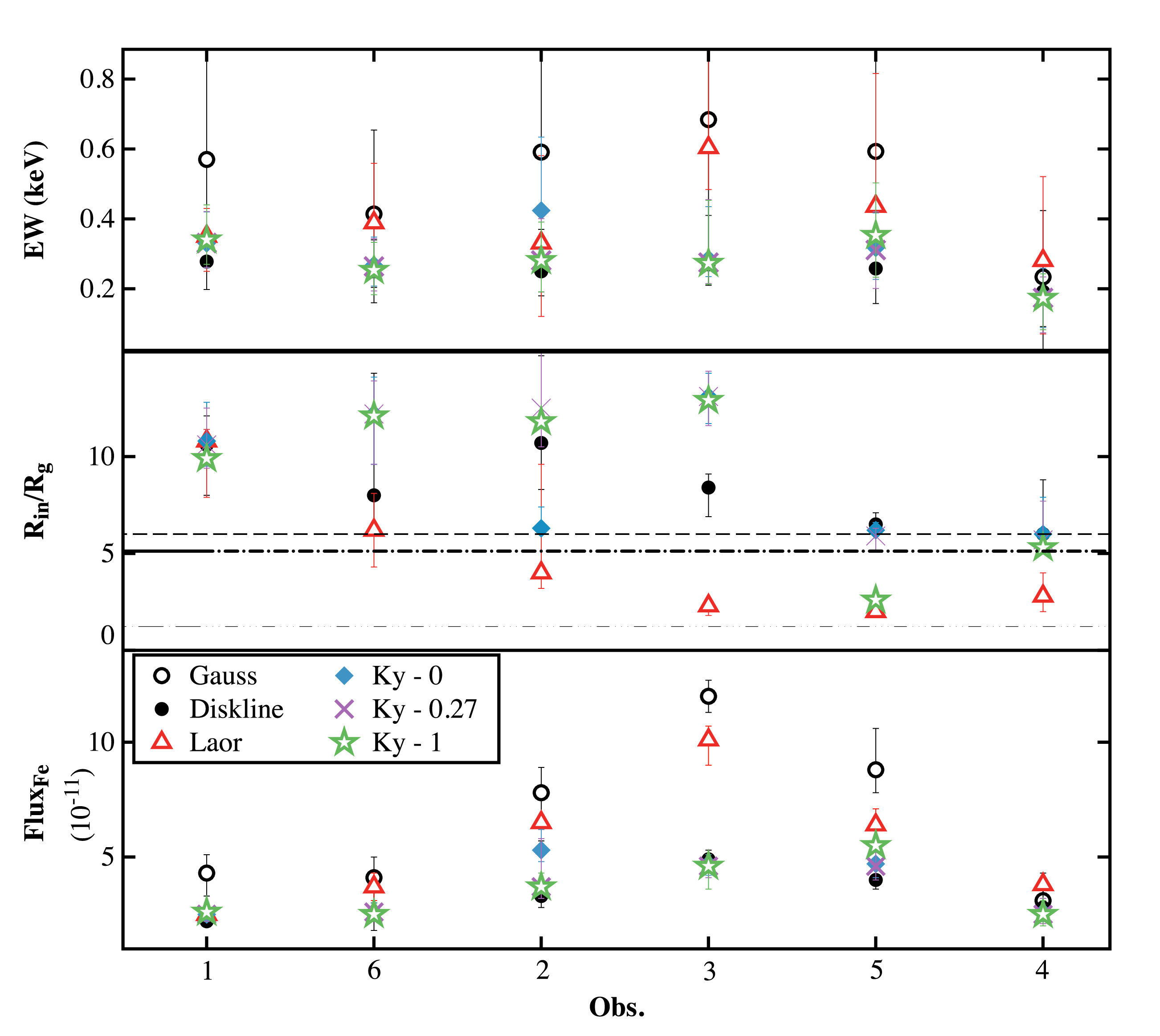}
\caption{Evolution of the Fe emission parameters of different line models as the source moves through the CD. The observations are ordered according to their $S_{\rm z}$ coordinate. From top to bottom, the panels show respectively the evolution of the equivalent width, the inner radius in units of the gravitational radius, $R_g=GM/c^2$, and the line unabsorbed flux in units of $10^{-11}$ erg~cm$^{-2}$~s$^{-1}$. Dashed, dashed-dot, and dashed-dot-dot lines (middle panel) represent the 6 $R_g$, 5.1 $R_g$, and 1.2 $R_g$ radii, respectively. }
\label{fig:line_par}
\end{center}
\end{figure*}

The first thing to notice from the table is the fact that the Fe line is very well modelled with a simple, symmetric, gaussian profile. Replacing the gaussian profile with a relativistic line model always lead to an increase of $\chi^2$ (with the exception of Obs. 4), with a maximum $\Delta \chi^2$ between the gaussian and the other models of $\sim$32 (for 2 d.o.f. difference). The gaussian component, however, showed always a very broad profile, with a line width between 1 and 1.4 keV that seemed to increase as the source went from the transitional-state to the soft state, with the exception of Obs.~4 in which the width dropped to its lowest value of 1 keV. In Obs.~1, 2, 5 and 6, all the model consistently showed an energy line at 6.4 keV that implies neutral or lowly ionised iron. In Obs.~3, \textsc{kyrline} gave energy values of $\sim$6.7 keV while the other models place the line at $\sim$6.4 keV. Finally, in Obs.~4 \textsc{diskline} and \textsc{kyrline} gave an energy value of $\sim$6.4 keV, while \textsc{gaussian} and \textsc{laor} gave values larger than 6.7 keV. The inclination was generally too high given that no dips or eclipses have been observed in 4U 1636--53. All the models showed inclinations larger than 70 degree, with most of the values showing an upper confidence limit that pegged at 90$^\circ$. We further found that the emissivity index of the disc was roughly the same for all observations for all the relativistic models. \\ 
In Figure~\ref{fig:line_par}, from top to bottom,  we show the evolution of the EW, inner radius and line flux across the CD for the different components used to model the Fe K-$\alpha$ line. \textsc{gaussian} and \textsc{laor} showed the largest EW, although with big errors. Within errors, we found the values of the EW to be consistent between the different models, and not varying with the source state. In the middle panel of Figure~\ref{fig:line_par} we show the inner radius in units of the gravitational radius, $R_g$. In the panel we plot 3 radii 
representing the ISCO for the Schwarzschild metric (6 $R_g$), the Kerr metric with $a_*=0.27$ ($\sim$5.1 $R_g$) and the Kerr metric with $a_*=1$ ($\sim$1.2 $R_g$). Interesting to notice is that only in Obs.~1 all the phenomenological models gave a consistent estimate of the inner radius of $\sim$ 11 $R_g$, while for the other observations the different models showed significantly different inner radii. Among the models, \textsc{laor} showed generally the lowest values of the radius. Although \textsc{laor} and \textsc{kyrline} with $a_*=1$ are based on  the same Kerr-metric kernel, in 4 out of 6 observations both gave significantly different values of $R_{in}$. A similar consideration is valid also for \textsc{diskline} and \textsc{kyrline} with $a_*=0$. Overall, the inner radius did not vary much from Obs.~6 to Obs.~3, and than significantly dropped in the last two observations.\\
Finally, the bottom panel of Figure~\ref{fig:line_par} shows the unabsorbed line flux in the 0.5--130 keV energy range. The \textsc{gaussian} and \textsc{laor} components showed the largest flux values in all observations. Furthermore, the flux values of both these line models increased from Obs.~6 to Obs.~3, and then decreased down to flux values comparable to the transitional-state ones. \textsc{diskline} and \textsc{kyrline} showed a more clear trend with $S\rm{z}$: the line fluxes increased from the transitional-state to the soft state with the only exception of Obs.~4 where the flux dropped.  

\begin{table}
\begin{center}
\begin{tabular}{cccc}\hline

\multicolumn{1}{c}{Obs.} & 
\multicolumn{1}{c}{\textsc{bbrefl+rfxconv}} & 
\multicolumn{1}{c}{\textsc{bbrefl}} &  
\multicolumn{1}{c}{\textsc{rfxconv}} \\\hline  
 
\textsc{1} & \cellcolor{Gray}313/278 & 326/279 & 334/279 \\
\textsc{2} & 286/278 & 286/279 &	 \cellcolor{Gray}288/279 \\
\textsc{3} & \cellcolor{Gray}285/278 & 	326/279 & 298/279\\
\textsc{4} & 301/278 &	\cellcolor{Gray}301/279 & 327/279\\ 
\textsc{5} & \cellcolor{Gray}287/268 &	302/269 & 308/269\\
\textsc{6} & 290/278 &	290/279  & \cellcolor{Gray}320/279\\\hline
\end{tabular}
\caption{$\chi^{2}$ and number of degrees of freedom for 3 different configurations for the incident emission. \textsc{bbrefl} models a reflection spectrum from a disc illuminated by a blackbody. \textsc{rfxconv} is used to create reflection assuming \textsc{nthcomp} as incident spectrum. The third option includes both contributions from \textsc{bbody} and \textsc{nthcomp} direct emissions. In grey we highlighted the best fit models chosen for the analysis. See Section~\ref{subsec:reflModel} for more details.}
\label{tab:modelsRefl}
\end{center}
\end{table}

\subsubsection{Emission line: phenomenological models vs. reflection}
As shown in section 3.4.2, the iron emission line is well fitted by a symmetric gaussian profile, characterised however by a large breadth. In order to explore the broadening mechanism that shapes the reflection spectrum, we started by fitting the reflection spectrum without the \textsc{kerrconv} kernel that models the relativistic effects. The model therefore was (see Sec.~\ref{sec:irradiation}): {\sc phabs*(diskbb + bbody + nthcomp + bbrefl + rfxconv*nthcomp)}. The only broadening mechanism in this case is Compton broadening. We found that in all the six observations the model did not fit well the data, with $\chi_{\nu}^{2}$ varying between 1.3 and 1.7 (with d.o.f. ranging from 271 to 282). From the residuals was clear a strong wiggled-like feature in the energy range 5 to 8 keV, consistent with the unfitted iron emission line. By adding the \textsc{kerrconv} component all the fits improved, as reported in Table~\ref{tab:reflection}.
We found that the inner radii inferred from the reflection model are consistent with the values from \textsc{kyrline} with $a_*=0.27$. Furthermore, the reflection model gave very large inclination values, generally larger than 80$^\circ$, consistent with \textsc{kyrline} (with the only exception of Obs. 2 in which the reflection model gave an inclination of $\sim$50$^\circ$). Finally, we noticed that the EW in the 4-9 keV energy range from the reflection model was consistent within errors with the EW from the relativistic model \textsc{kyrline}.


%
%
%

\section{Discussion}\label{sec:discussion}

We fitted the X-ray spectra of six simultaneous RXTE and XMM-Newton observations of the NS low-mass X-ray binary 4U 1636-53. For the first time we used a reflection model in which simultaneously both the NS surface/boundary layer and the corona were taken to be the source of irradiation to investigate the relative reflection of each of them. We further modelled the Fe emission line assuming different line profiles, both symmetric (\textsc{gaussian}) and relativistically-broadened (\textsc{diskline}, \textsc{laor}, and  \textsc{kyrline}).

We found that, both in the transitional state and the soft state, the corona is the most prominent component in the broadband spectrum. The emission from the NS surface/boundary layer is on average contributing an order of magnitude less than the corona to the continuum emission. However, the relative contribution of the NS surface/boundary layer increases going from the transitional to the soft state. In four out of six observations (Obs.~1 in the transitional state, and Obs.~3, 4 and 5 in the soft state) the NS surface/boundary layer is the main source of disc irradiation, accounting for $\sim$70\% and $\sim$80\% of the reflected spectrum in the Fe line region (based on the 4--9 keV unabsorbed flux) in the transitional and soft state, respectively. The NS surface/boundary layer reflection spectrum is even more predominant in Obs.~4, where the corona does not contribute at all to the reflection. This may be due to the NS surface/boundary layer being the main contributor to the reflection spectrum simply because is closer to the inner edge of the accretion disc where most of the disc emission generates. On the other hand is also likely that part of the photons emitted by the NS surface/boundary layer interact with the corona by means of Compton scattering, as a consequence of that what we model as direct emission from the NS is only an underestimation.\\
For the other two observations (Obs.~2 in the soft state, and Obs.~6 in the transitional state) the reflection spectrum originates entirely from the corona emission. 
\citet{Cackett10} analysed Obs.~1--3 using a blurred reflection model where only the NS surface/boundary layer (fitted with a blackbody component) was assumed to illuminate the disc. We found that the line properties inferred from our fits with the reflection model are consistent within errors with \citet{Cackett10} findings. The only discrepancy between the results is related to the logarithm of the ionisation parameter for Obs.~1, which we find to be $\sim$1 while \citet{Cackett10} find an ionisation parameter of $\sim$2.5. It is not surprising, at least for Obs.~1 and 3, that our results matched \citet{Cackett10} findings. As mentioned above, in Obs.~1 and 3, the NS surface/boundary layer is the predominant source of disc irradiation in the Fe line region, accounting for 70\% and 85\% of the reflected spectrum, respectively. On the other hand, it is interesting to mention that although in Obs.~2 we modelled the reflection spectrum assuming the corona as source of irradiation, the line parameters are still consistent with those of \citet{Cackett10}.

 We modelled the Fe emission line assuming different line profiles, both symmetric (\textsc{gaussian}) and relativistically-broadened (\textsc{diskline}, \textsc{laor}, and  \textsc{kyrline}). We found that all the models fit reasonably well the data (see Table~\ref{tab:simple_line} for all the fitting parameters). The \textsc{gaussian} profile seemed to fit the line better, which suggests a symmetric profile for the Fe emission line. This result agrees with \citet{Ng10}, that found the lines from Obs.~1--3 to be symmetric, and that the \textsc{gaussian} and the relativistic profile \textsc{laor} could fit equally well the line. \citet{Ng10} suggested that, contrary to previous claims, the width of the line can be explained by mechanisms other than relativistic effects, such as Compton broadening. \citet{Kallman89} showed that iron lines generated in the ADC are characterised by a symmetric profile due to blending, Compton scattering and rotation. \citet{Kallman89} further estimated that Fe lines generated in the ADC could show EW values up to 100 eV assuming standard parameters. From the fits with the \textsc{gaussian} model we found EW values between $\sim$300 and $\sim$700 eV, which are significantly larger than \citet{Kallman89} estimates. This indicates that Compton scattering alone cannot explain the EW from the \textsc{gaussian} model. This is also confirmed by the reflection model: We find that excluding the convolution component \textsc{kerrconv}, the reflection model fails to fit the Fe line in the spectrum. For completeness we should mention that the EWs reported by \citet{Ng10} are significantly lower (between $\sim$30 and $\sim$200 eV) than the ones we report here, and that \citet{Ng10} raised doubts on the Fe line with the largest EW (Obs.~1) being realistic. The fact that we use a wide bandpass energy spectrum (0.8--120 keV) to constrain the direct emission explains the differences between our results and those of \citet{Ng10}. That said, our results suggest that even though the Fe emission line profile is compatible with being symmetric, the broad profile requires a broadening mechanism other than Compton broadening, and therefore the relativistically-broadened interpretation is still compatible with the data. 

The inner disc radii measured from the relativistic models range between 2 and 13 $R_g$. We noticed some differences between models based on different space-time metrics. \textsc{laor} (for maximally rotating black holes) systematically showed the lowest values of $R_{in}$, in the range 2 to 6 $R_g$ (with the exception of Obs.~1 where all the models measured $R_{in}\sim$11 $R_g$). \textsc{diskline}, based on a Schwarzschild metric, showed $R_{in}$ between 6 and 11 $R_g$, with one observation pegging to the lowest value. Between the two models, \textsc{laor} fits the data statistically better in most of the observations. Notice, however, that for rotating NS with $a_*<0.3$ the metric is expected not to deviate much from Schwarzschild metric \citep[see][for more details]{Miller98}, which is probably the reason why \textsc{laor} under-estimated the inner disc radius.
\citet{Cackett10} fitted the emission line profile on Obs.~1--3 using the \textsc{diskline} model. It is interesting to notice that the measurement for  the inner radius and the EW from \citet{Cackett10} are lower than what we reported here. On the other hand, line energy and emission index are consistent with being the same. The fact that \citet{Cackett10} did not apply corrections for pileup effects and that only used the XMM-Newton data to fit the Fe line, may be the reasons of these discrepancies with our results.

Finally, we modelled the Fe line profile with \textsc{kyrline} that allows to set the spin parameter between 0 (equivalent to a Schwarzschild metric) and 1 (maximally rotating Kerr black hole). We tested three different spin parameter values, $a_*=$ 0, 0.27 and 1 (see Section~\ref{subsec:direct} for details). With these models we measured values of the inner disc radius mostly between 5 and 13 $R_g$ (with the only exception of Obs.~5 where for \textsc{kyrline} with $a_*=$1 we found the radius to be at $\sim$2.5 $R_g$). We found consistent measurements of the inner radius for all three spin parameters, with only two exception in Obs.~2 and 5, where one of the models significantly deviated from the other two (see Table~\ref{tab:simple_line}).
It is interesting to mention that \textsc{kyrline} with $a_*=$1 and \textsc{laor}, that are based on the same Kerr metric showed in general significantly different values for the Fe line parameters \citep[see][for a detail comparison between the models]{Svoboda09}. Very similar to that, also for \textsc{kyrline} with $a_*=$0 and \textsc{diskline} we noticed discrepancies in the modelling of the line profile, although less significant.

None of the line parameters from the relativistic models show a clear trend with the source state. However, the inner radius of the disc appears to significantly change for two of the soft-state observations (Obs.~4 and 5), the two with the largest $S_z$ values. The line flux increased going from the transitional-state to the soft-state.
The inner radius inferred from the reflection model, which is consistent with $R_{in}$ from \textsc{kyrline} with $a_*=$0.27, did not show any clear correlation with the source state. 

Our findings, both from reflection and phenomenological models, suggest that at least for 4U 1636--53 the properties of the Fe emission line, assuming the line profile is broadened by relativistic effects, are not consistent with the scenario in which the inner truncation radius of the accretion disc changes with the source state.

\subsection{Caveats of the analysis}
\label{sec:caveats}
In all cases, the model we used to fit the six \textit{XMM-Newton/RXTE} observations of 4U 1636--53, including relativistically smeared ionised reflection, described the data well, although there are a few caveats. Firstly, we generally found ionisation parameter values consistent with the presence of moderately or highly ionised Fe ($\log \xi$ between 1.2 and 3), however the line energy values from the phenomenological models were mostly at 6.4 keV consistent with neutral Fe. This has been also observed by \citet{Cackett10} between the reflection model and \textsc{diskline}.  The combination of the thermal emission from the disc and the incident emission are consistent with ionisation parameters we found, raising then doubts on the reliability of the line energy values from the phenomenological models. However, we should keep in mind that with phenomenological models we oversimplify the reflection spectrum between 4-9 keV by only fitting a emission line while in reality the spectrum is a superposition of a continuum plus a line spectrum. 

Secondly, from all our fits to 4U 1636--53, we found a high disc inclination, which is at odds with the fact that no eclipses are detected in this source. From previous analyses of the first three observations in our sample, \citet{Pandel08} and \citet{Cackett10} also reported similar inclinations, always with an upper confidence limit pegged at 90$^{\circ}$. The high inclination in the analysis of \citet{Pandel08} and \citet{Cackett10} maybe partly due to pileup effects, as pileup hardens the spectrum and possibly extends the blue wing of the Fe line. \citet{Pandel08} also suggested that the broad line could be due to line blending of different ionisation stages of Fe. When they fit the Fe line with a multiple line model, they found the inclination to be slightly lower ($>64^{\circ}$), although it was still consistent with 90$^{\circ}$. Nevertheless, even with a conservative pileup correction, and fitting a model that self-consistently includes line blending effects and Compton broadening \citep[the latter was not included in the model used by][]{Pandel08}, we still found very high inclination values. Although fits to PN timing mode data of other NS LMXBs have revealed well constrained and much lower inclination angles \citep[see e.g.][]{DAi09,Iaria09}, it is possible that the PN timing mode data we have used here are still affected by calibration issues. It would be useful to perform a spectral analysis similar to the one presented in this work, but applying it to data taken in different data modes or with other X-ray missions. 
%


%
\section{Summary}\label{sec:summary}
We analysed six \textit{XMM-Newton/RXTE} observations of the NS LMXB 4U 1636--53; in two of the observations the source was in the transitional state, and in the other four the source was in the soft state. We used a relativistically smeared ionised reflection model to study the Fe emission line apparent in all six spectra, which we used to investigate the relative contribution of the corona and the NS surface/boundary layer to the reflection continuum. We found that the NS surface/boundary is the main source of irradiation in the Fe line region in four out six observations, both in the transitional (also known as island state) and soft state. In the other two observations (one in the transitional and the other in the soft state), the whole reflected spectrum is due to the corona illuminating the accretion disc.

\noindent
We also fitted the Fe emission line with a set of phenomenological models including symmetric as well as relativistically-broadened profiles. We found that, although the relativistic line models fitted well the data, the symmetric \textsc{gaussian} profile gave the best fit in statistical terms. The best-fitting Gaussian line was, however, in all cases very broad. Our results suggested that such a broad line profile is unlikely to be produced by Compton broadening, and that relativistic broadening may still be necessary.
We found that relativistic models did not consistently modelled the Fe line; this is especially true for the \textsc{laor} model. Values of the inner radius estimated form the reflection models are consistent with the results from \textsc{kyrline} with $a_*=0.27$. However,  the reflection model did not help solving the issue of the too high inclination values derived from the Fe line profile. 

\noindent
Finally, we also explored the variation of the direct continuum emission and the properties of the Fe line as a function of the source state. According to the standard accretion disc model, as mass accretion rate increases the disc moves inwards. We found that the continuum emission overall matched consistently with that picture, however the inner disc radius inferred from the Fe line did not show any significant correlation with the source state.

%
\section{Acknowledgments}
We are grateful to Chris Done for providing the \textsc{rfxconv} model and for useful comments and discussion regarding the model and the manuscript. AS and MM wish to thank ISSI for their hospitality. TB acknowledges support from ASI-INAF grant I/009/10/0.
%
%
%

\clearpage
\begin{table*}
\begin{minipage}{180mm}
\begin{center}{
\scriptsize \caption{Obs. 1-3 -- Phenomenological models parameters: Continuum components}\label{tab:simple_cont1}
\begin{tabular}{cclcccccc}\hline

\multicolumn{1}{c}{Obs. 1} & 
\multicolumn{1}{c}{Model comp} & 
\multicolumn{1}{l}{Parameter} & 
\multicolumn{1}{c}{\textsc{Gaussian}} & 
\multicolumn{1}{c}{\textsc{Diskline}} & 
\multicolumn{1}{c}{\textsc{Laor}} & 
\multicolumn{1}{c}{\textsc{Kyrline 0}}&
\multicolumn{1}{c}{\textsc{Kyrline 0.27}}&
\multicolumn{1}{c}{\textsc{Kyrline 1}}\\\hline
\\
& \textsc{phabs}   & $N_{\rm H}$ ($10^{22}$ cm$^{-2}$)   & 0.42$\pm$0.08  & 0.34$^{+0.05}_{-0.09}$ & 0.41$^{+0.11}_{-0.10}$          & 0.41$^{+0.06}_{-0.11}$ &  0.40$^{+0.07}_{-0.10}$ & 0.47$^{+0.05}_{-0.04}$ \\
& \textsc{diskbb}  & $kT_{\rm in}$ (keV)    & 0.19$^{+0.03}_{-0.02}$ & 0.18$\pm$0.06 & 0.17$^{+0.03}_{-0.01}$  & 0.18$^{+0.04}_{-0.02}$ & 0.18$^{+0.03}_{-0.02}$ & 0.20$^{+0.02}_{-0.01}$ \\
&                  & $N_{\rm dbb}$ & 16648$^{+15985}_{-9988}$ & 11238$^{+18530}_{-995}$ & 27086$^{+18564}_{-19156}$ & 17387$^{+17190}_{-11320}$ & 18114$^{+23283}_{-11337}$ & 10498$^{+7584}_{-5862}$\\
&                 & F$_{\rm d}$ ($10^{-11}$)     & 14.9$^{+23.0}_{-11.3}$   & 7.4$^{+30.2}_{-6.8}$         & 12.9$^{+21.4}_{-10.1}$       & 11.5$^{+25.6}_{-10.0}$      & 12.0$^{+22.0}_{-9.9}$       & 12.5$^{+12.2}_{-7.6}$\\\\
& \textsc{bbody}  & $kT_{\rm BB}$ (keV)  & 1.30$^{+0.15}_{-0.06}$ & 1.56$^{+0.05}_{-0.06}$ & 1.56$^{+0.11}_{-0.08}$ & 1.56$\pm$0.07 & 1.56$\pm$0.07 & 1.56$\pm$0.08\\
&                  &$N_{\rm BB}$ ($10^{-3}$)& 0.9$\pm$0.2 & 1.33$\pm$0.2  & 1.1$\pm$0.2 & 1.1$\pm$0.2 & 1.1$\pm$0.1 & 1.1$\pm$0.1\\
&                 & F$_{\rm b}$ ($10^{-10}$)    & 0.8$\pm$0.2  & 1.2$\pm$0.1       &0.9$\pm$0.2        & 0.9$\pm$0.2        & 0.9$\pm$0.1      & 0.9$\pm$0.2\\\\
& \textsc{nthcomp}& $\Gamma$   & 1.85$\pm$0.02 & 1.84$\pm$0.06 & 1.85$\pm$0.01 & 1.85$\pm$0.01 & 1.85$\pm$0.01 & 1.86$\pm$0.01 \\
&				&  $kT_{\rm e}$ (keV) & 15.4$^{+1.3}_{-1.0}$ & 15.2$^{+1.1}_{-0.9}$ & 15.4$^{+1.3}_{-1.1}$ & 15.4$^{+1.1}_{-0.9}$ & 15.3$^{+1.1}_{-1.0}$ & 15.5$^{+1.2}_{-1.0}$  \\
&                  & $N_{\rm NTH}$  & 0.198$^{+0.008}_{-0.007}$ & 0.193$^{+0.004}_{-0.008}$ & 0.201$^{+0.007}_{-0.008}$ & 0.199$^{+0.004}_{-0.007}$ & 0.198$^{+0.005}_{-0.002}$ & 0.200$\pm$0.004\\
 &                & F$_{\rm NTH}$ ($10^{-9}$)    & 2.1$\pm$0.1  &    2.0$^{+0.3}_{-0.4}$     & 2.1$\pm$0.1       & 2.0$^{+0.2}_{-0.1}$          & 2.0$^{+0.8}_{-0.6}$        & 2.0$\pm$0.7\\\\
&                  & $\chi^{2}_{\nu}$ ($\chi^{2}/\nu$) & 1.08 (306/284) & 1.18 (333/282) & 1.09 (307/282) & 1.11 (312/282) & 1.11 (312/282) & 1.12 (317/282)  \\\\
&           & Total flux ($10^{-9}$)        & 2.2$^{+0.3}_{-0.2}$ & 2.2$\pm$0.4       & 2.2$^{+0.2}_{-0.1}$      & 2.2$^{+0.3}_{-0.1}$       & 2.2$^{+0.2}_{-0.1}$       & 2.2$\pm$0.1\\\\

\hline


\multicolumn{1}{c}{Obs. 2} & 
\multicolumn{1}{c}{Model comp} & 
\multicolumn{1}{l}{Parameter} & 
\multicolumn{1}{c}{\textsc{Gaussian}} & 
\multicolumn{1}{c}{\textsc{Diskline}} & 
\multicolumn{1}{c}{\textsc{Laor}} & 
\multicolumn{1}{c}{\textsc{Kyrline 0}}&
\multicolumn{1}{c}{\textsc{Kyrline 0.27}}&
\multicolumn{1}{c}{\textsc{Kyrline 1}}\\\hline
\\
& \textsc{phabs}   & $N_{\rm H}$ ($10^{22}$ cm$^{-2}$)    & 0.30$\pm$0.05  & 0.32$\pm$0.02 & 0.31$\pm$0.03  & 0.31$\pm$0.03 &  0.31$\pm$0.03 & 0.32$^{+0.02}_{-0.03}$ \\
& \textsc{diskbb}  & $kT_{\rm in}$ (keV)  & 0.65$^{+0.03}_{-0.08}$ & 0.67$\pm$0.02 & 0.65$\pm$0.04  & 0.68$\pm$0.04 & 0.67$^{+0.02}_{-0.04}$ & 0.67$^{+0.05}_{-0.03}$\\
&                  & $N_{\rm dbb}$ & 107$^{+29}_{-15}$ & 188$^{+24}_{-19}$ & 127$^{+22}_{-14}$ & 159$^{+22}_{-23}$ & 172$^{+28}_{-23}$ & 173$^{+18}_{-23}$\\
&                 & F$_{\rm d}$ ($10^{-11}$)     & 33.3$^{+11.5}_{-15.6}$   & 66.7$^{+12.3}_{-11.7}$         & 39.6$^{+13.4}_{-10.4}$       & 60.1$^{+18.6}_{-16.3}$      & 61.0$^{+12.9}_{-14.0}$       & 61.4$^{+22.9}_{-13.6}$\\\\
& \textsc{bbody}  & $kT_{\rm BB}$ (keV)  & 1.76$\pm$0.05 & 1.64$\pm$0.03 & 1.75$\pm$0.05 & 1.68$\pm$0.05 & 1.68$^{+0.04}_{-0.05}$ & 1.68$^{+0.04}_{-0.03}$\\
&                  &$N_{\rm BB}$ ($10^{-3}$)& 3.8$^{+0.8}_{-0.5}$ & 6.1$^{+0.6}_{-0.3}$  & 4.4$^{+0.1}_{-0.2}$ & 5.3$\pm$0.7 & 5.8$^{+0.6}_{-0.5}$ & 5.8$^{+0.6}_{-0.5}$\\
&                 & F$_{\rm b}$ ($10^{-10}$)    & 3.2$^{+0.7}_{-0.4}$  & 5.1$^{+0.5}_{-0.2}$      &3.7$\pm$0.2        & 4.5$\pm$0.6        & 4.9$\pm$0.5      & 4.9$\pm$0.5\\\\
& \textsc{nthcomp}& $\Gamma$   & 2.3$\pm$0.1 & 2.1$^{+0.1}_{-0.2}$ & 2.3$\pm$0.1 & 2.2$\pm$0.2 & 2.2$^{+0.2}_{-0.3}$ & 2.2$^{+0.2}_{-0.1}$ \\
&				&  $kT_{\rm e}$ (keV) & 6.1$^{+0.8}_{-0.3}$ & 5.4$\pm$0.2 & 6.2$^{+0.3}_{-0.2}$ & 5.6$^{+1.4}_{-0.4}$ & 5.6$^{+0.3}_{-0.5}$ & 5.6$^{+0.9}_{-0.5}$  \\
&                  & $N_{\rm NTH}$  & 0.29$^{+0.06}_{-0.02}$ & 0.19$^{+0.01}_{-0.02}$ & 0.27$^{+0.06}_{-0.03}$ & 0.21$^{+0.02}_{-0.06}$ & 0.21$^{+0.03}_{-0.05}$ & 0.21$^{+0.04}_{-0.03}$\\
 &                & F$_{\rm NTH}$ ($10^{-9}$)    & 2.1$^{+0.4}_{-0.3}$  &    1.7$^{+0.2}_{-0.5}$     & 2.0$^{+0.5}_{-0.3}$       & 1.7$^{+0.3}_{-0.5}$          & 1.7$^{+0.4}_{-0.8}$        & 1.7$^{+0.4}_{-0.3}$\\\\
&                  & $\chi^{2}_{\nu}$ ($\chi^{2}/\nu$) & 1.05 (298/284) & 1.07 (304/282) & 1.06 (299/282) & 1.07 (301/282) & 1.07 (303/282) & 1.07 (302/282)  \\\\
&           & Total flux ($10^{-9}$)        & 2.8$^{+0.5}_{-0.4}$ & 2.8$^{+0.2}_{-0.6}$       & 2.8$^{+0.5}_{-0.3}$      & 2.8$^{+0.4}_{-0.7}$       & 2.8$^{+0.4}_{-0.8}$       & 2.8$^{+0.5}_{-0.3}$\\\\
\hline
\multicolumn{1}{c}{Obs. 3} & 
\multicolumn{1}{c}{Model comp} & 
\multicolumn{1}{l}{Parameter} & 
\multicolumn{1}{c}{\textsc{Gaussian}} & 
\multicolumn{1}{c}{\textsc{Diskline}} & 
\multicolumn{1}{c}{\textsc{Laor}} & 
\multicolumn{1}{c}{\textsc{Kyrline 0}}&
\multicolumn{1}{c}{\textsc{Kyrline 0.27}}&
\multicolumn{1}{c}{\textsc{Kyrline 1}}\\\hline
\\
& \textsc{phabs}   & $N_{\rm H}$ ($10^{22}$ cm$^{-2}$)    & 0.30$\pm$0.01  & 0.36$^{+0.04}_{-0.08}$ & 0.36$\pm$0.01  & 0.36$\pm$0.01 &  0.37$\pm$0.09 & 0.36$^{+0.02}_{-0.03}$ \\
& \textsc{diskbb}  & $kT_{\rm in}$ (keV)  & 0.79$^{+0.02}_{-0.08}$  & 0.73$\pm$0.01 & 0.71$\pm$0.03  & 0.73$^{+0.03}_{-0.01}$  & 0.73$^{+0.03}_{-0.01}$ & 0.73$^{+0.05}_{-0.03}$\\
&                  & $N_{\rm dbb}$ & 162$^{+70}_{-3}$ & 237$^{+3}_{-4}$ & 244$\pm$51 & 245$^{+5}_{-3}$ & 253$\pm$38 & 244$^{+18}_{-23}$\\
&                 & F$_{\rm d}$ ($10^{-9}$)     & 1.2$^{+0.5}_{-0.4}$   & 1.2$\pm$0.1         & 1.1$\pm$0.3      & 1.2$^{+0.2}_{-0.1}$      & 1.3$^{+0.3}_{-0.2}$       & 1.2$^{+0.4}_{-0.2}$\\\\
& \textsc{bbody}  & $kT_{\rm BB}$ (keV)  & 1.43$^{+0.10}_{-0.43}$ & 1.48$\pm$0.03 & 1.33$\pm$0.07 & 1.46$^{+0.06}_{-0.03}$ & 1.46$^{+0.05}_{-0.03}$ & 1.46$^{+0.04}_{-0.03}$\\
&                  &$N_{\rm BB}$ ($10^{-3}$)& 6.9$\pm$0.2 & 9.9$^{+0.1}_{-0.4}$  & 8.2$^{+0.5}_{-2.1}$ & 9.8$^{+0.1}_{-1.0}$ & 10.4$^{+0.1}_{-0.1}$ & 10.3$^{+0.6}_{-0.5}$\\
&                 & F$_{\rm b}$ ($10^{-10}$)    & 5.8$\pm$1.7  & 8.3$^{+0.1}_{-0.3}$      &6.9$^{+0.4}_{-1.8}$        & 8.3$^{+0.1}_{-0.8}$        & 8.8$\pm$0.8      & 8.7$^{+0.5}_{-2.6}$\\\\
& \textsc{nthcomp}& $\Gamma$   & 1.85$^{+0.05}_{-0.02}$ & 1.82$^{+0.03}_{-0.02}$ & 1.82$^{+0.01}_{-0.02}$ & 1.80$^{+0.10}_{-0.03}$ & 1.7$^{+0.2}_{-0.1}$ & 1.8$^{+0.2}_{-0.1}$ \\
&				&  $kT_{\rm e}$ (keV) & 3.3$\pm$0.1  & 3.4$\pm$0.1 & 3.2$\pm$0.1 & 3.3$\pm$0.1 & 3.3$\pm$0.1 & 3.3$^{+0.9}_{-0.5}$  \\
&                  & $N_{\rm NTH}$  & 0.15$\pm$0.01 & 0.13$^{+0.04}_{-0.02}$ & 0.15$^{+0.03}_{-0.01}$ & 0.13$\pm$0.01 & 0.11$\pm$0.01 & 0.11$^{+0.04}_{-0.03}$\\
&                 & F$_{\rm NTH}$ ($10^{-9}$)    & 1.7$\pm$0.2 &    1.5$^{+0.5}_{-0.2}$     & 1.6$^{+0.3}_{-0.1}$       & 1.5$^{+0.2}_{-0.1}$          & 1.5$^{+0.4}_{-0.3}$        & 1.3$^{+0.6}_{-0.4}$\\\\
&                  & $\chi^{2}_{\nu}$ ($\chi^{2}/\nu$) & 1.09 (306/281) & 1.19 (333/279) & 1.12 (314/279) & 1.16 (325/279) & 1.17 (326/279) & 1.17 (328/279)  \\\\
&           & Total flux ($10^{-9}$)        & 3.4$\pm$0.5 & 3.5$^{+0.5}_{-0.2}$       & 3.5$^{+0.5}_{-0.4}$      & 3.5$^{+0.3}_{-0.2}$       & 3.5$^{+0.5}_{-0.4}$       & 3.5$^{+0.7}_{-0.5}$\\\\
\hline
\hline
\end{tabular}
\normalsize}
\end{center}
NOTES.--~A * means that the error was pegged at the hard limit of the parameter range. All uncertainties are given at 90\% confidence level. $N_{\rm dbb}$ is defined
as $(R_{in}/D_{10})^2\cos \theta$, with $R_{in}$ in km, $D_{10}$ the distance in 10 kpc, and $\theta$ the inclination angle of the disc. $N_{\rm BB}$ is $L_{39}/D_{10}^2$, where $L_{39}$
is the luminosity in units of $10^{39}$ erg s$^{-1}$. $N_{\rm NTH}$ is in units of photons keV$^{-1}$ cm$^{-2}$ s$^{-1}$ at 1 keV. All fluxes represent the unabsorbed 0.5--130 keV flux in
units of erg~cm$^{-2}$~s$^{-1}$.
\end{minipage}
\end{table*}

\clearpage

\begin{table*}
\begin{minipage}{180mm}
\begin{center}{
\scriptsize \caption{Obs. 4-6 -- Phenomenological models parameters: Continuum components}\label{tab:simple_cont2}
\begin{tabular}{cclcccccc}\hline

\multicolumn{1}{c}{Obs. 4} & 
\multicolumn{1}{c}{Model comp} & 
\multicolumn{1}{l}{Parameter} & 
\multicolumn{1}{c}{\textsc{Gaussian}} & 
\multicolumn{1}{c}{\textsc{Diskline}} & 
\multicolumn{1}{c}{\textsc{Laor}} & 
\multicolumn{1}{c}{\textsc{Kyrline 0}}&
\multicolumn{1}{c}{\textsc{Kyrline 0.27}}&
\multicolumn{1}{c}{\textsc{Kyrline 1}}\\\hline
\\
& \textsc{phabs}   & $N_{\rm H}$ ($10^{22}$ cm$^{-2}$)    & 0.30$\pm$0.01  & 0.30$\pm$0.01 & 0.31$\pm$0.02  & 0.30$\pm$0.03 &  0.30$\pm$0.01 & 0.30$\pm$0.01 \\
& \textsc{diskbb}  & $kT_{\rm in}$ (keV)  & 0.76$\pm$0.03 & 0.77$\pm$0.02 & 0.74$\pm$0.02  & 0.78$\pm$0.02 & 0.78$^{+0.04}_{-0.01}$ & 0.78$\pm$0.01\\
&                  & $N_{\rm dbb}$ & 149$^{+11}_{-13}$ & 142$^{+7}_{-4}$ & 136$^{+10}_{-13}$ & 150$^{+8}_{-4}$ & 153$^{+8}_{-6}$ & 151$\pm$8\\
&                 & F$_{\rm d}$ ($10^{-10}$)     & 3.4$^{+0.8}_{-0.7}$   & 3.5$^{+0.6}_{-0.5}$         & 2.7$\pm$0.5      & 4.0$^{+0.6}_{-0.5}$      & 4.1$^{+1.3}_{-0.3}$       & 4.0$\pm$0.4 \\\\
& \textsc{bbody}  & $kT_{\rm BB}$ (keV)  & 1.67$\pm$0.03 & 1.73$^{+0.02}_{-0.08}$ & 1.72$\pm$0.04 & 1.70$\pm$0.02 & 1.70$\pm$0.02 & 1.71$\pm$0.02\\
&                  &$N_{\rm BB}$ ($10^{-3}$)& 8.6$^{+1.6}_{-2.6}$ & 9.3$\pm$1.3  & 7.8$\pm$0.9 & 9.6$\pm$1.6 & 10.1$\pm$1.3 & 10.1$^{+1.3}_{-3.1}$\\
&                 & F$_{\rm b}$ ($10^{-10}$)    & 5.8$^{+1.1}_{-1.8}$  & 6.2$\pm$0.9      &5.2$^{+0.6}_{-0.7}$        & 6.4$\pm$1.1       & 6.8$\pm$0.9      & 6.7$^{+0.9}_{-2.1}$\\\\
& \textsc{nthcomp}& $\Gamma$   & 1.9$^{+0.3}_{-0.1}$ & 2.0$\pm$0.2  & 2.1$^{+0.4}_{-0.2}$ & 1.9$^{+0.4}_{-0.2}$ & 1.8$\pm$0.3 & 1.8$^{+0.3}_{-0.2}$ \\
&				&  $kT_{\rm e}$ (keV) & 3.8$^{+1.3}_{-0.2}$ & 4.2$\pm$0.2 & 4.1$\pm$0.2 & 4.0$\pm$0.2 & 3.9$^{+0.3}_{-0.1}$ & 4.0$^{+1.7}_{-0.2}$  \\
&                  & $N_{\rm NTH}$  & 0.10$\pm$0.02 & 0.09$\pm$0.01 & 0.14$^{+0.03}_{-0.02}$ & 0.07$\pm$0.01 & 0.06$\pm$0.01 & 0.06$\pm$0.02\\
&                 & F$_{\rm NTH}$ ($10^{-10}$)    & 6.1$^{+1.9}_{-1.4}$&    5.0$^{+0.9}_{-1.3}$     & 6.9$^{+2.3}_{-1.8}$       & 4.3$^{+1.4}_{-1.3}$          & 4.2$^{+1.3}_{-2.4}$        & 4.2$^{+1.7}_{-1.3}$\\\\
&                  & $\chi^{2}_{\nu}$ ($\chi^{2}/\nu$) & 1.08 (304/281) & 1.06 (297/279) & 1.09 (304/279) & 1.06 (295/279) & 1.06 (296/279) & 1.06 (296/279)  \\\\
&           & Total flux ($10^{-9}$)        & 1.5$^{+0.2}_{-0.3}$ & 1.5$^{+0.1}_{-0.2}$       & 1.5$\pm$0.2      & 1.5$\pm$0.2       & 1.5$^{+0.2}_{-0.3}$       & 1.5$^{+0.2}_{-0.3}$\\\\
\hline


\multicolumn{1}{c}{Obs. 5} & 
\multicolumn{1}{c}{Model comp} & 
\multicolumn{1}{l}{Parameter} & 
\multicolumn{1}{c}{\textsc{Gaussian}} & 
\multicolumn{1}{c}{\textsc{Diskline}} & 
\multicolumn{1}{c}{\textsc{Laor}} & 
\multicolumn{1}{c}{\textsc{Kyrline 0}}&
\multicolumn{1}{c}{\textsc{Kyrline 0.27}}&
\multicolumn{1}{c}{\textsc{Kyrline 1}}\\\hline
\\
& \textsc{phabs}   & $N_{\rm H}$ ($10^{22}$ cm$^{-2}$)    & 0.28$\pm$0.02  & 0.30$^{+0.01}_{-0.03}$ & 0.29$^{+0.05}_{-0.01}$  & 0.29$\pm$0.02 &  0.29$\pm$0.01 & 0.29$\pm$0.01 \\
& \textsc{diskbb}  & $kT_{\rm in}$ (keV)  & 0.75$\pm$0.04 & 0.75$\pm$0.01 & 0.73$\pm$0.02  & 0.79$\pm$0.02 & 0.79$^{+0.01}_{-0.02}$ & 0.79$^{+0.05}_{-0.03}$\\
&                  & $N_{\rm dbb}$ & 71$^{+19}_{-29}$ & 146$^{+7}_{-10}$ & 120$^{+14}_{-9}$ & 146$^{+11}_{-7}$ & 154$^{+10}_{-8}$ & 148$^{+11}_{-15}$\\
&                 & F$_{\rm d}$ ($10^{-10}$)     & 1.5$^{+0.6}_{-0.7}$   & 3.1$\pm$0.3         & 6.1$^{+1.0}_{-0.8}$      & 10.4$^{+1.4}_{-1.2}$      & 10.1$\pm$1.0       & 10.5$^{+3.2}_{-1.9}$\\\\
& \textsc{bbody}  & $kT_{\rm BB}$ (keV)  & 1.97$\pm$0.13 & 1.68$\pm$0.02 & 1.67$^{+0.04}_{-0.17}$ & 1.63$\pm$0.03 & 1.63$\pm$0.02 & 1.64$\pm$0.03\\
&                  &$N_{\rm BB}$ ($10^{-3}$)& 4.9$^{+0.6}_{-0.4}$ & 6.7$^{+1.6}_{-0.2}$  & 4.4$\pm$0.7 & 7.4$\pm$1.1 & 7.7$\pm$1.3 & 7.4$^{+4.0}_{-0.4}$\\
&                 & F$_{\rm b}$ ($10^{-10}$)    & 3.0$^{+0.4}_{-0.3}$  & 4.5$^{+1.0}_{-0.1}$      &3.7$\pm$0.6       & 6.2$\pm$0.9       & 6.5$\pm$1.1     & 6.2$^{+3.3}_{-0.3}$\\\\
& \textsc{nthcomp}& $\Gamma$   & 2.4$\pm$0.2 & 2.0$\pm$0.1  & 2.1$\pm$0.1 & 1.9$^{+0.1}_{-0.2}$ & 1.8$\pm$0.1 & 1.9$\pm$0.1 \\
&				&  $kT_{\rm e}$ (keV) & 4.2$^{+0.9}_{-0.7}$ & 3.6$^{+0.1}_{-0.8}$ & 3.4$^{+0.8}_{-0.1}$ & 3.5$^{+0.2}_{-0.7}$ & 3.4$\pm$0.1 & 3.5$\pm$0.1  \\
&                  & $N_{\rm NTH}$  & 0.29$^{+0.11}_{-0.04}$ & 0.18$^{+0.04}_{-0.02}$ & 0.23$^{+0.01}_{-0.03}$ & 0.12$\pm$0.01 & 0.11$^{+0.03}_{-0.01}$ & 0.12$\pm$0.01\\
&                 & F$_{\rm NTH}$ ($10^{-9}$)    & 1.1$^{+0.5}_{-0.3}$&    1.0$\pm$0.2     & 1.9$^{+0.2}_{-0.3}$       & 1.3$^{+0.2}_{-0.4}$          & 1.3$^{+0.5}_{-0.2}$        & 1.3$\pm$0.2\\\\
&                  & $\chi^{2}_{\nu}$ ($\chi^{2}/\nu$) & 1.12 (304/272) & 1.22 (331/270) & 1.15 (310/270) & 1.17 (317/270) & 1.18 (320/270) & 1.21 (327/270)  \\\\
&           & Total flux ($10^{-9}$)        & 3.0$^{+0.5}_{-0.3}$ & 3.0$^{+0.3}_{-0.2}$       & 3.0$^{+0.2}_{-0.3}$       & 3.0$^{+0.2}_{-0.5}$      & 3.0$^{+0.5}_{-0.3}$       & 3.0$^{+0.5}_{-0.3}$\\\\
\hline

\multicolumn{1}{c}{Obs. 6} & 
\multicolumn{1}{c}{Model comp} & 
\multicolumn{1}{l}{Parameter} & 
\multicolumn{1}{c}{\textsc{Gaussian}} & 
\multicolumn{1}{c}{\textsc{Diskline}} & 
\multicolumn{1}{c}{\textsc{Laor}} & 
\multicolumn{1}{c}{\textsc{Kyrline 0}}&
\multicolumn{1}{c}{\textsc{Kyrline 0.27}}&
\multicolumn{1}{c}{\textsc{Kyrline 1}}\\\hline
\\
& \textsc{phabs}   & $N_{\rm H}$ ($10^{22}$ cm$^{-2}$)    & 0.32$\pm$0.05  & 0.27$^{+0.03}_{-0.09}$ & 0.32$^{+0.05}_{-0.09}$  & 0.31$^{+0.03}_{-0.06}$ &  0.31$^{+0.04}_{-0.08}$ & 0.29$^{+0.06}_{-0.04}$ \\
& \textsc{diskbb}  & $kT_{\rm in}$ (keV)  & 0.32$\pm$0.05 & 0.41$^{+0.04}_{-0.02}$ & 0.33$^{+0.06}_{-0.05}$  & 0.34$^{+0.06}_{-0.03}$ & 0.35$^{+0.06}_{-0.03}$ & 0.37$\pm$0.05\\
&  & $N_{\rm dbb}$ & 713$^{+466}_{-251}$ & 360$^{+187}_{-28}$ & 644$^{+435}_{-124}$ & 611$^{+217}_{-54}$ & 581$^{+422}_{-114}$ & 466$^{+599}_{-69}$\\
&                 & F$_{\rm d}$ ($10^{-10}$)     & 0.9$^{+1.1}_{-0.6}$   & 1.5$^{+1.1}_{-0.4}$         & 1.0$^{+1.3}_{-0.6}$      & 1.1$^{+1.2}_{-0.4}$      & 1.1$^{+1.2}_{-0.2}$       & 1.2$^{+2.0}_{-0.3}$\\\\
& \textsc{bbody}  & $kT_{\rm BB}$ (keV)  & 1.79$\pm$0.13 & 1.73$^{+0.04}_{-0.12}$ & 1.90$\pm$0.11 & 1.89$\pm$0.08 & 1.88$\pm$0.02 & 1.86$\pm$0.08\\
&  &$N_{\rm BB}$ ($10^{-3}$)& 1.7$\pm$0.3 & 2.5$^{+0.3}_{-0.2}$  & 1.8$\pm$0.2 & 2.2$\pm$0.1 & 2.2$\pm$0.2 & 2.2$\pm$0.2\\
&                 & F$_{\rm b}$ ($10^{-10}$)    & 1.4$\pm$0.2  & 2.1$^{+0.3}_{-0.6}$      &1.5$\pm$0.2      & 1.9$\pm$0.2       & 1.9$\pm$0.2     & 1.9$\pm$0.2\\\\
& \textsc{nthcomp}& $\Gamma$   & 1.9$\pm$0.1 & 1.9$\pm$0.1  & 1.9$^{+0.2}_{-0.1}$ & 1.9$\pm$0.1 & 1.9$\pm$0.1 & 1.9$\pm$0.1  \\
&	&  $kT_{\rm e}$ (keV) & 15.7$^{+4.2}_{-2.6}$ & 14.5$\pm$1.8 & 16.8$^{+5.0}_{-2.8}$& 16.8$^{+4.2}_{-3.0}$ & 16.5$^{+4.5}_{-2.7}$ & 16.2$^{+3.8}_{-2.7}$  \\
&  & $N_{\rm NTH}$  & 0.23$\pm$0.02 & 0.19$\pm$0.01 & 0.23$\pm$0.02 & 0.22$\pm$0.02 & 0.22$^{+0.01}_{-0.03}$ & 0.21$\pm$0.01\\
&                 & F$_{\rm NTH}$ ($10^{-9}$)    & 2.3$^{+0.5}_{-0.7}$&    2.1$^{+0.4}_{-0.6}$     & 2.4$^{+0.8}_{-0.7}$       & 2.3$^{+0.5}_{-0.7}$          & 2.3$^{+0.5}_{-0.7}$        & 2.3$^{+0.4}_{-0.6}$\\\\
& & $\chi^{2}_{\nu}$ ($\chi^{2}/\nu$) & 0.93 (259/278) & 1.06 (291/276) & 0.94 (260/276) & 0.97 (268/276) & 0.97 (268/276) & 0.98 (270/276)\\\\
&           & Total flux ($10^{-9}$)        & 2.6$^{+0.5}_{-0.7}$ & 2.6$^{+0.4}_{-0.5}$       & 2.6$^{+0.8}_{-0.7}$       & 2.6$^{+0.5}_{-0.7}$      & 2.6$^{+0.5}_{-0.7}$       & 2.6$^{+0.5}_{-0.6}$\\\\
\hline
\end{tabular}
\normalsize}
\end{center}
NOTES.--~A * means that the error was pegged at the hard limit of the parameter range. All uncertainties are given at 90\% confidence level. $N_{\rm dbb}$ is defined
as $(R_{in}/D_{10})^2\cos \theta$, with $R_{in}$ in km, $D_{10}$ the distance in 10 kpc, and $\theta$ the inclination angle of the disc. $N_{\rm BB}$ is $L_{39}/D_{10}^2$, where $L_{39}$
is the luminosity in units of $10^{39}$ erg s$^{-1}$. $N_{\rm NTH}$ is in units of photons keV$^{-1}$ cm$^{-2}$ s$^{-1}$ at 1 keV. All fluxes represent the unabsorbed 0.5--130 keV flux in
units of erg~cm$^{-2}$~s$^{-1}$.
\end{minipage}
\end{table*}

\clearpage
\begin{table*}
\begin{minipage}{180mm}
\begin{center}{
\scriptsize \caption{Phenomenological model parameters: Fe line}\label{tab:simple_line}
\resizebox{0.93\columnwidth}{!}{
\begin{tabular}{ccccccccccc}\hline
\multicolumn{1}{c}{} & 
\multicolumn{1}{c}{Model} & 
\multicolumn{1}{c}{E$_{\rm line}$ (keV)} & 
\multicolumn{1}{c}{ $\sigma$ (keV)} & 
\multicolumn{1}{c}{incl\,($^\circ$)} & 
\multicolumn{1}{c}{$R_{\rm in}/R_{g}$} & 
\multicolumn{1}{c}{$\beta$}&
\multicolumn{1}{c}{Norm ($10^{-3}$)}&
\multicolumn{1}{c}{Flux ($10^{-11}$)}&
\multicolumn{1}{c}{EW (keV)}&
\multicolumn{1}{c}{$\chi^2/dof$}\\\hline
\\
 &\textsc{gauss}& 6.40$^{+0.07}_{-0.0*}$&1.27$^{+0.10}_{-0.14}$ & -- &-- & -- &4.2$^{+0.8}_{-1.0}$ &4.3$^{+0.8}_{-1.0}$ &0.57$^{+0.36}_{-0.31}$ & 306/284\\\\
 &\textsc{diskline}&6.40$^{+0.02}_{-0.0*}$&  --& 90.0$^{+0.0*}_{-15.9}$&  10.6$^{+1.5}_{-2.6}$& -2.7$\pm$0.1&  2.1$\pm$0.2& 2.2$\pm$0.2&  0.28$^{+0.06}_{-0.08}$ &333/282\\\\
 &\textsc{laor}&6.44$^{+0.08}_{-0.04*}$& --& 86.3$^{+0.1}_{-0.3}$& 10.8$^{+0.6}_{-2.9}$& 4.4$^{+1.9}_{-0.8}$& 2.5$^{+0.4}_{-0.3}$& 2.5$^{+0.4}_{-0.3}$&0.35$^{+0.08}_{-0.10}$&307/282\\\\
\textbf{Obs. 1} &\textsc{Ky} a$_*$=0& 6.40$^{+0.06}_{-0.0*}$& --& 86.1$^{+0.7}_{-0.9}$& 10.8$^{+2.0}_{-1.3}$& 3.3$^{+0.4}_{-0.3}$& 2.4$\pm$0.3& 2.5$^{+0.3}_{-0.5}$& 0.33$^{+0.09}_{-0.07}$&312/282\\\\
 &\textsc{Ky} a$_*$=0.27& 6.40$^{+0.06}_{-0.0*}$& -- & 85.9$^{+0.7}_{-0.9}$& 10.6$^{+1.9}_{-1.2}$& 3.3$^{+0.2}_{-0.3}$& 2.4$\pm$0.3&  2.5$^{+0.4}_{-0.3}$&0.33$^{+0.09}_{-0.07}$&312/282\\\\
 &\textsc{Ky} a$_*$=1& 6.40$^{+0.06}_{-0.0*}$& --& 85.1$^{+0.8}_{-0.9}$& 9.9$^{+1.9}_{-1.1}$& 3.1$^{+0.3}_{-0.2}$& 2.5$\pm$0.3& 2.6$\pm$0.3& 0.34$^{+0.10}_{-0.07}$ &317/282 \\\\\hline 
 &\textsc{gauss}& 6.40$^{+0.07}_{-0.0*}$& 1.3$\pm$0.1& -- & -- & --& 7.6$^{+1.1}_{-2.0}$& 7.8$^{+1.1}_{-2.1}$ &0.59$^{+0.43}_{-0.26}$ &298/284\\\\
 &\textsc{diskline}& 6.40$^{+0.07}_{-0.0*}$& --& 72.5$^{+17.5*}_{-11.7}$	& 10.7$^{+4.5}_{-2.4}$& -2.7$\pm$0.2& 3.2$\pm$0.5& 3.3$\pm$0.5&  0.25$^{+0.12}_{-0.07}$&304/282	\\\\
 &\textsc{laor}& 6.40$^{+0.24}_{-0.0*}$& --& 86.8$^{+3.2*}_{-0.4}$& 4.0$^{+5.6}_{-0.8}$& 4.0$^{+1.4}_{-0.7}$& 6.4$\pm$0.1	& 6.5$\pm$0.3& 0.33$^{+0.25}_{-0.21}$&299/282	\\\\
 \textbf{Obs. 2} &\textsc{Ky} a$_*$=0& 6.40$^{+0.2}_{-0.0*}$& --& 88.9$^{+1.1*}_{-0.9}$& 6.3$^{+1.1}_{-0.3*}$& 3.1$^{+0.5}_{-0.4}$& 5.3$^{+0.9}_{-0.7}$& 5.3$^{+0.9}_{-0.7}$& 0.42$^{+0.21}_{-0.11}$	&301/282	\\\\
 &\textsc{Ky} a$_*$=0.27& 6.40$^{+0.1}_{-0.0*}$& --& 83.9$^{+2.1}_{-7.9}$& 12.5$^{+2.9}_{-2.0}$& 3.0$^{+0.7}_{-0.4}$& 3.6$^{+2.0}_{-0.6}$& 3.7$^{+2.1}_{-0.5}$& 0.28$^{+0.12}_{-0.09}$&303/282 \\\\
 &\textsc{Ky} a$_*$=1&6.40$^{+0.1}_{-0.0*}$& --& 83.0$^{+2.1}_{-5.8}$& 11.8$^{+3.2}_{-1.7}$	& 2.9$^{+0.5}_{-0.3}$& 3.6$^{+0.6}_{-0.3}$& 3.7$^{+0.6}_{-0.3}$& 0.28$^{+0.11}_{-0.09}$&302/282\\\\\hline
 &\textsc{gauss}& 6.40$^{+0.06}_{-0.0*}$	 & 1.4$\pm$0.1& --& --& --	& 11.7$\pm$0.7& 12.0$\pm$0.7&  0.68$^{+0.51}_{-0.42}$&306/281\\\\
 &\textsc{diskline}&6.40$^{+0.03}_{-0.0*}$& --& 90.0$^{+0.0*}_{-15.9}$& 8.4$^{+0.7}_{-1.5}$& -2.7$\pm$0.8& 4.6$\pm$0.4& 4.9$^{+0.4}_{-0.5}$&  0.27$^{+0.14}_{-0.06}$&333/279\\\\
 &\textsc{laor}&6.40$^{+0.05}_{-0.0*}$& --& 87.5$^{+2.5*}_{-0.5}$& 2.3$^{+0.2}_{-0.5}$& 3.5$\pm$0.2& 9.6$^{+0.5}_{-1.0}$& 10.1$^{+0.6}_{-1.1}$& 0.60$^{+0.55}_{-0.12}$&314/279\\\\
 \textbf{Obs. 3} &\textsc{Ky} a$_*$=0&6.67$^{+0.06}_{-0.07}$& --& 86.2$^{+0.6}_{-0.8}$& 13.1$^{+1.2}_{-1.4}$& 3.6$^{+0.5}_{-0.3}$& 4.4$\pm$0.4	& 4.7$^{+0.4}_{-0.5}$& 0.29$^{+0.15}_{-0.05}$&325/279\\\\
 &\textsc{Ky} a$_*$=0.27&6.68$\pm$0.07& --& 86.1$^{+0.7}_{-0.9}$& 13.1$^{+1.3}_{-1.5}$& 3.5$^{+0.4}_{-0.3}$& 4.3$\pm$0.4& 4.6$^{+0.4}_{-0.5}$& 0.28$^{+0.18}_{-0.06}$&326/279\\\\
 &\textsc{Ky} a$_*$=1&6.70$\pm$0.1& --	& 85.6$^{+2.1}_{-5.8}$& 12.9$^{+3.2}_{-1.7}$& 3.4$^{+0.5}_{-0.3}$&4.3$^{+0.6}_{-0.3}$& 4.6$^{+0.1}_{-1.0}$& 0.27$^{+0.18}_{-0.06}$&328/279\\\\\hline
  &\textsc{gauss}&6.86$^{+0.11*}_{-0.10}$ & 1.0$\pm$0.2& --	& -- & --& 2.9$^{+1.1}_{-0.8}$& 3.1$^{+1.2}_{-0.9}$&  0.23$^{+0.19}_{-0.21}$ &304/281 	\\\\
 &\textsc{diskline}& 6.43$^{+0.05}_{-0.03*}$ & --& 73.4$^{+4.6}_{-5.9}$& 6.0$^{+2.8}_{-0.0*}$& -2.6$\pm$0.1& 2.6$^{+1.0}_{-0.4}$& 2.8$^{+1.0}_{-0.4}$&  0.19$\pm$0.10&297/279 \\\\
 &\textsc{laor}&	6.72$^{+0.15}_{-0.32*}$& --& 88.1$^{+1.9*}_{-1.8}$& 2.8$^{+1.2}_{-0.8}$& 3.3$^{+0.9}_{-0.3}$& 3.5$\pm$0.1& 3.8$^{+0.1}_{-0.2}$& 0.28$^{+0.24}_{-0.21}$&304/279		\\\\
 \textbf{Obs. 4} &\textsc{Ky} a$_*$=0&6.43$^{+0.1}_{-0.03*}$& -- & 72.3$^{+9.2}_{-4.9}$& 6.0$^{+1.9}_{-0.0*}$& 2.4$^{+0.2}_{-0.1}$& 2.4$^{+0.6}_{-0.5}$& 2.6$\pm$0.6 & 0.18$^{+0.08}_{-0.09}$&295/279\\\\
 &\textsc{Ky} a$_*$=0.27&6.43$^{+0.07}_{-0.03*}$& --& 72.1$^{+9.2}_{-5.2}$& 5.7$^{+2.0}_{-0.6*}$& 2.4$^{+0.2}_{-0.1}$& 2.4$^{+0.6}_{-0.4}$& 2.5$^{+0.7}_{-0.4}$& 0.17$^{+0.06}_{-0.10}$&296/279 \\\\
 &\textsc{Ky} a$_*$=1&6.44$^{+0.07}_{-0.04*}$& --& 74.3$^{+7.3}_{-5.6}$& 5.3$^{+1.4}_{-1.9}$& 2.4$^{+0.2}_{-0.1}$& 2.3$^{+0.6}_{-0.4}$& 2.5$^{+0.6}_{-0.5}$& 0.17$^{+0.07}_{-0.09}$&296/279		\\\\\hline
  &\textsc{gauss}&6.41$^{+0.09}_{-0.01*}$& 1.4$\pm$0.1& -- & --& --& 8.6$^{+1.8}_{-1.0}$& 8.8$^{+1.8}_{-1.0}$&  0.59$^{+0.29}_{-0.23}$&304/272\\\\
 &\textsc{diskline}&6.40$^{+0.02}_{-0.0*}$ & -- & 89.9$^{-89.9*}_{-14.5}$& 6.5$^{+0.6}_{-0.5*}$& -2.6$\pm$0.1 & 3.8$^{+0.6}_{-0.4}$& 4.0$^{+0.5}_{-0.4}$&  0.29$\pm$0.10&331/270 \\\\
 &\textsc{laor}&6.40$^{+0.06}_{-0.0*}$& --& 90.0$^{-0.0*}_{-1.7}$& 2.0$^{+0.4}_{-0.2}$& 3.4$\pm$0.1& 6.1$\pm$0.7& 6.4$\pm$0.7& 0.44$^{+0.38}_{-0.12}$&310/270\\\\
 \textbf{Obs. 5} &\textsc{Ky} a$_*$=0&6.40$^{+0.04}_{-0.0*}$& --& 87.4$^{+0.9}_{-1.0}$& 6.2$^{+0.4}_{-0.2*}$& 2.8$\pm$0.2& 4.6$\pm$0.5& 4.7$^{+0.1}_{-0.5}$& 0.32$^{+0.10}_{-0.9}$&317/270\\\\
 &\textsc{Ky} a$_*$=0.27&6.40$^{+0.04}_{-0.0*}$& --& 86.7$\pm$1.5& 5.9$^{+0.4}_{-0.8*}$ & 2.7$^{+0.2}_{-0.1}$& 4.5$^{+0.6}_{-0.5}$& 4.6$\pm$0.6& 0.31$\pm$0.11&320/270\\\\
 &\textsc{Ky} a$_*$=1&6.40$^{+0.03}_{-0.0*}$& --& 89.0$^{+1.0*}_{-1.3}$& 2.6$\pm$0.1& 2.4$\pm$0.2& 5.4$\pm$0.6& 5.5$\pm$0.2& 0.35$^{+0.15}_{-0.12}$&327/270	\\\\\hline
  &\textsc{gauss}&6.40$^{+0.05}_{-0.0*}$	& 1.2$\pm$0.1& --& --& --& 4.0$\pm$1.0& 4.1$^{+0.9}_{-1.1}$&  0.41$^{+0.24}_{-0.21}$&259/278	\\\\
 &\textsc{diskline}& 6.40$^{+0.02}_{-0.0*}$& --	& 90.0$^{+0.0*}_{-21.5}$& 8.0$^{+6.3}_{-2.0*}$& -2.5$\pm$0.1& 2.4$^{+0.3}_{-0.7}$& 2.5$^{+0.3}_{-0.7}$&  0.25$\pm$0.09&291/276 	\\\\
 &\textsc{laor}&6.40$^{+0.07}_{-0.0*}$ & --& 86.4$^{+0.3}_{-0.1}$& 6.2$\pm$1.9& 3.8$^{+1.3}_{-0.5}$& 3.7$\pm$0.6& 3.7$\pm$0.6& 0.39$^{+0.17}_{-0.12}$ &260/276\\\\
 \textbf{Obs. 6} &\textsc{Ky} a$_*$=0&6.40$^{+0.07}_{-0.0*}$ & -- & 86.8$^{+0.7}_{-0.9}$& 12.2$^{+1.9}_{-2.6}$ & 3.3$^{+0.6}_{-0.4}$& 2.6$^{+0.3}_{-0.2}$& 2.6$^{+0.4}_{-0.2}$& 0.27$^{+0.08}_{-0.06}$&268/276  \\\\
 &\textsc{Ky} a$_*$=0.27&6.40$^{+0.07}_{-0.0*}$& -- & 86.5$^{+0.7}_{-0.9}$& 12.2$^{+1.7}_{-2.6}$& 3.3$^{+0.6}_{-0.4}$	& 2.5$\pm$0.3& 2.6$\pm$0.3& 0.26$^{+0.08}_{-0.07}$&268/276 \\\\
 &\textsc{Ky} a$_*$=1&6.40$^{+0.09}_{-0.0*}$	& --& 85.8$^{+0.9}_{-1.0}$& 12.1$^{+2.0}_{-2.1}$& 3.1$^{+0.5}_{-0.3}$& 2.5$\pm$0.3& 2.5$\pm$0.3& 0.25$^{+0.08}_{-0.07}$&270/276 	\\\\\hline
\end{tabular}}
\normalsize}
\end{center}
NOTES.--~A * means that the error was pegged at the hard limit of the parameter range. All uncertainties are given at 90\% confidence level. $\beta$ represents the power law dependence of the disc emissivity.
The normalisation of all the models is the line flux in photons cm$^{-2}$ s$^{-1}$. Line flux represents the unabsorbed 0.5--130 keV flux in units of erg~cm$^{-2}$~s$^{-1}$. Inner radius is in units of gravitational radius $Rg=(GM/c^2)$. 
\end{minipage}
\end{table*}

\clearpage

\appendix

\section{Pileup correction for PN data}\label{app:pile-up}

\begin{figure}
\begin{center}
\includegraphics[width=3.2in]{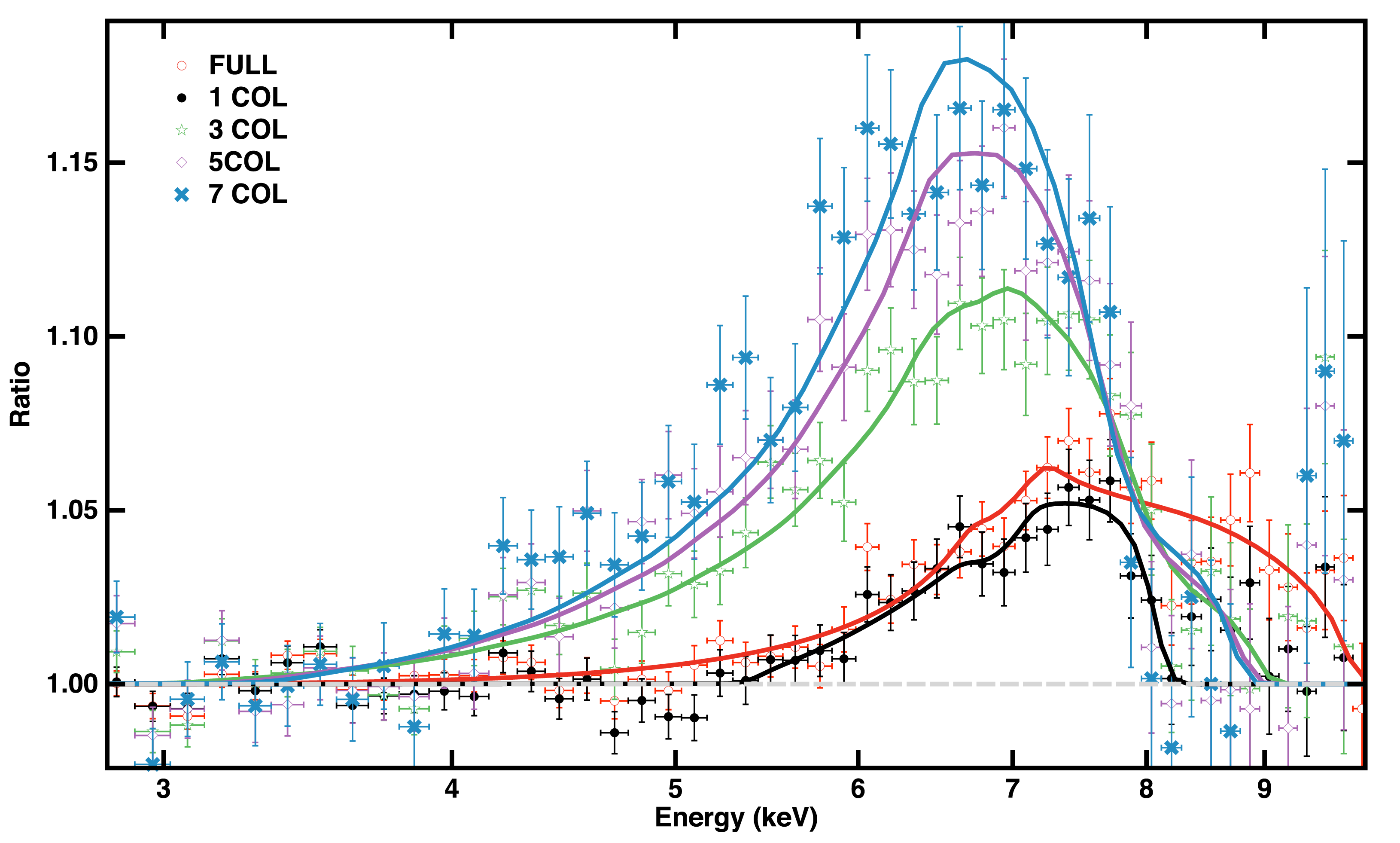} 
\end{center}
\caption{Data/model ratios in the 3--9 keV energy range, for different extraction regions used to generate the PN spectra for Obs.~2. Different colour points represent data from different extraction regions, while coloured-solid lines represent the best-fit line profiles. We simultaneously fitted six PN spectra, including also the simultaneous PCA and HEXTE data. We modelled the Fe emission line using the relativistic line model \textsc{kyrline} with $a_*= $ 0.27.  The residuals are represented after setting the line normalisation to 0.}
\label{fig:pileup}
\end{figure}

As reported by, e.g., \citet{Ng10}, data can suffer from pileup even when the average PN count rate is below the critical pileup level reported by the XMM-Newton User handbook (800 cts~s$^{-1}$). Triggered by that we inspected the dataset for pileup effects. In addition to the spectra created for the full 17-column wide box centred on the source position, we extracted spectra selecting the same events as before but now excising the central, 1, 3, 5 and 7 columns, respectively. We then used the task \texttt{epatplot} to determine whether each of these spectra have significant pileup. For Obs. 1 (with a count rate of $\sim$250 cts s$^{-1}$), we found that excising only the single central column was sufficient to significantly reduce the pileup. For the other five ($>$ 300 cts s$^{-1}$) observations, we had to excise the 3 central columns to correct for pileup. Unfortunately, the task \texttt{epatplot} does not provide a quantitative measure of how pileup affects the spectrum, nor whether the Fe emission line suffers from that. To investigate the effect of pileup on the Fe emission line, we fitted the spectra of all datasets, including the simultaneous PCA and HEXTE data, using a phenomenological model for the Fe line (for practical reasons we will only discuss results about Obs.~2, that represents the soft state). We fitted simultaneously all the spectra previously used for the \texttt{epatplot} investigation, where we excised 0, 1, 3, 5 and 7 columns. We left the continuum spectral shape parameters (see Section~\ref{subsec:direct} for details on the model) free to vary, but we linked them across different spectra, we also included the simultaneous PCA and HEXTE data. By doing that we forced the continuum emission to maintain the same spectral shape. We only left the normalisation of all the continuum components free to vary in all the spectra. We modelled the Fe emission line with the relativistic model \textsc{kyrline} assuming a$_*$=0.27. The line parameters were free in all spectra. The main assumption here is that if the data are not affected by pileup, the continuum spectral shape will not vary while excising central columns from the PSF. This, in principle, should also apply to the Fe line profile. In Figure~\ref{fig:pileup} we show the data/model ratio in the energy range 3--9 keV, after setting the normalisation of the iron line to 0. Different colour represent different excising regions, going from none (black points) to 7 column excised (light-blue points), and solid lines represent the best-fit Fe line profile. From Figure~\ref{fig:pileup} we notice that the line profile significantly changes with the number of excised columns. More specifically, we find that the line energy quickly drops from 6.97 to 6.4 keV as the number of columns excised goes from 0 to 3. On the other hand excising 5 and 7 columns still gives line energies around 6.4 keV. The inner radius shows a similar behaviour, as we can notice from the line profile. Figure~\ref{fig:pileup} shows that pileup strongly affects the Fe line. From these tests, we decided for this specific observation to correct for pileup by excising 5 central columns of the PSF.
Summarising, based on the \texttt{epatplot} task and this test, we decided to excise 3 central columns for Obs.~1, Obs. 4--6, and we excised 5 columns for Obs.~2 and Obs.~3. Our extraction regions for Obs.~1--3 are more conservative in comparison with the regions used in previous analysis of the same data \citep[see][]{Ng10}.

To further investigate whether pileup affected our data, we fitted the
full spectra (no columns excised) and the spectra where we excised a
number of the central columns, leaving all fitting parameters of the
model (the same model used for the previous test) free to vary. We found
that in all six observations the best-fitting parameters of some of the
spectral components, including the iron line, changed significantly as
we excised more and more central columns of the PSF, up to a point in
which the changes were no longer significant when we continued excising
columns. As an example, for Obs.~2, we found that the blackbody emission
from the NS surface/boundary layer and the iron line components changed
significantly when we fit the full data (no columns excised) and the
data where we excised the central 5 columns. More specifically, the
temperature of the blackbody decreases from $1.96 \pm 0.02$ keV when we
fitted the spectrum of the full PSF, to $1.68 \pm 0.05$ keV when we
excised the central 5 columns. At the same time the inner radius of the
disc inferred from the iron line changed from $28^{+19}_{-11}$ $R_g$ to
$12.5 \pm 2.5$ $R_g$, the line energy from $6.97^{+0}_{-0.05}$ keV to
$6.4^{+0.1}_{-0}$ keV, and the line normalisation from $0.6 \pm 0.1$
$10^{-3}$ photons cm$^{-2}$ s$^{-1}$ to $3.6 \pm 0.4$ $10^{-3}$ photons
cm$^{-2}$ s$^{-1}$. Both the decrease of the black body temperature, and
the changes of the iron line parameters while excising the central
columns of the point spread function are consistent with pileup effects.

\newpage


\begin{thebibliography}{}

\bibitem[\protect\citeauthoryear{{Altamirano}, {van der Klis}, {M{\'e}ndez},
  {Jonker}, {Klein-Wolt} \& {Lewin}}{{Altamirano} et~al.}{2008}]{Altamirano08}
{Altamirano} D.,  {van der Klis} M.,  {M{\'e}ndez} M.,  {Jonker} P.~G.,
  {Klein-Wolt} M.,    {Lewin} W.~H.~G.,  2008, \apj, 685, 436

\bibitem[\protect\citeauthoryear{{Arnaud}}{{Arnaud}}{1996}]{Arnaud96}
{Arnaud} K.~A.,  1996, in {G.~H.~Jacoby \& J.~Barnes} ed., Astronomical Data
  Analysis Software and Systems V Vol.~101 of Astronomical Society of the
  Pacific Conference Series, {XSPEC: The First Ten Years}.
p.~17

\bibitem[\protect\citeauthoryear{{Ballantyne}}{{Ballantyne}}{2004}]{Ballantyne%
04}
{Ballantyne} D.~R.,  2004, \mnras, 351, 57

\bibitem[\protect\citeauthoryear{{Barret}, {Olive}, {Boirin}, {Done}, {Skinner}
  \& {Grindlay}}{{Barret} et~al.}{2000}]{Barret00}
{Barret} D.,  {Olive} J.~F.,  {Boirin} L.,  {Done} C.,  {Skinner} G.~K.,
  {Grindlay} J.~E.,  2000, \apj, 533, 329

\bibitem[\protect\citeauthoryear{{Belloni}, {Homan}, {Motta}, {Ratti} \&
  {M{\'e}ndez}}{{Belloni} et~al.}{2007}]{Belloni07}
{Belloni} T.,  {Homan} J.,  {Motta} S.,  {Ratti} E.,    {M{\'e}ndez} M.,  2007,
  \mnras, 379, 247

\bibitem[\protect\citeauthoryear{{Bhattacharyya} \&
  {Strohmayer}}{{Bhattacharyya} \& {Strohmayer}}{2007}]{Bhattacharyya07}
{Bhattacharyya} S.,  {Strohmayer} T.~E.,  2007, \apjl, 664, L103

\bibitem[\protect\citeauthoryear{{Braje}, {Romani} \& {Rauch}}{{Braje}
  et~al.}{2000}]{Braje00}
{Braje} T.~M.,  {Romani} R.~W.,    {Rauch} K.~P.,  2000, \apj, 531, 447

\bibitem[\protect\citeauthoryear{{Brenneman} \& {Reynolds}}{{Brenneman} \&
  {Reynolds}}{2006}]{Brenneman06}
{Brenneman} L.~W.,  {Reynolds} C.~S.,  2006, \apj, 652, 1028

\bibitem[\protect\citeauthoryear{{Cackett}, {Miller}, {Ballantyne}, {Barret},
  {Bhattacharyya}, {Boutelier}, {Miller}, {Strohmayer} \& {Wijnands}}{{Cackett}
  et~al.}{2010}]{Cackett10}
{Cackett} E.~M.,  {Miller} J.~M.,  {Ballantyne} D.~R.,  {Barret} D.,
  {Bhattacharyya} S.,  {Boutelier} M.,  {Miller} M.~C.,  {Strohmayer} T.~E.,
  {Wijnands} R.,  2010, \apj, 720, 205

\bibitem[\protect\citeauthoryear{{Cackett}, {Miller}, {Bhattacharyya},
  {Grindlay}, {Homan}, {van der Klis}, {Miller}, {Strohmayer} \&
  {Wijnands}}{{Cackett} et~al.}{2008}]{Cackett08}
{Cackett} E.~M.,  {Miller} J.~M.,  {Bhattacharyya} S.,  {Grindlay} J.~E.,
  {Homan} J.,  {van der Klis} M.,  {Miller} M.~C.,  {Strohmayer} T.~E.,
  {Wijnands} R.,  2008, \apj, 674, 415

\bibitem[\protect\citeauthoryear{{Casares}, {Cornelisse}, {Steeghs}, {Charles},
  {Hynes}, {O'Brien} \& {Strohmayer}}{{Casares} et~al.}{2006}]{Casares06}
{Casares} J.,  {Cornelisse} R.,  {Steeghs} D.,  {Charles} P.~A.,  {Hynes}
  R.~I.,  {O'Brien} K.,    {Strohmayer} T.~E.,  2006, \mnras, 373, 1235

\bibitem[\protect\citeauthoryear{{D'A{\`i}}, {di Salvo}, {Ballantyne}, {Iaria},
  {Robba}, {Papitto}, {Riggio}, {Burderi}, {Piraino}, {Santangelo}, {Matt},
  {Dov{\v c}iak} \& {Karas}}{{D'A{\`i}} et~al.}{2010}]{DAi10}
{D'A{\`i}} A.,  {di Salvo} T.,  {Ballantyne} D.,  {Iaria} R.,  {Robba} N.~R.,
  {Papitto} A.,  {Riggio} A.,  {Burderi} L.,  {Piraino} S.,  {Santangelo} A.,
  {Matt} G.,  {Dov{\v c}iak} M.,    {Karas} V.,  2010, \aap, 516, A36

\bibitem[\protect\citeauthoryear{{D'A{\`i}}, {Iaria}, {Di Salvo}, {Matt} \&
  {Robba}}{{D'A{\`i}} et~al.}{2009}]{DAi09}
{D'A{\`i}} A.,  {Iaria} R.,  {Di Salvo} T.,  {Matt} G.,    {Robba} N.~R.,
  2009, \apjl, 693, L1

\bibitem[\protect\citeauthoryear{{di Salvo}, {D'A{\'{\i}}}, {Iaria}, {Burderi},
  {Dov{\v c}iak}, {Karas}, {Matt}, {Papitto}, {Piraino}, {Riggio}, {Robba} \&
  {Santangelo}}{{di Salvo} et~al.}{2009}]{diSalvo09}
{di Salvo} T.,  {D'A{\'{\i}}} A.,  {Iaria} R.,  {Burderi} L.,  {Dov{\v c}iak}
  M.,  {Karas} V.,  {Matt} G.,  {Papitto} A.,  {Piraino} S.,  {Riggio} A.,
  {Robba} N.~R.,    {Santangelo} A.,  2009, \mnras, 398, 2022

\bibitem[\protect\citeauthoryear{{Done} \& {Gierli{\'n}ski}}{{Done} \&
  {Gierli{\'n}ski}}{2006}]{Done06}
{Done} C.,  {Gierli{\'n}ski} M.,  2006, \mnras, 367, 659

\bibitem[\protect\citeauthoryear{{Done}, {Gierli{\'n}ski} \& {Kubota}}{{Done}
  et~al.}{2007}]{Done07}
{Done} C.,  {Gierli{\'n}ski} M.,    {Kubota} A.,  2007, \aapr, 15, 1

\bibitem[\protect\citeauthoryear{{Dov{\v c}iak}, {Karas} \& {Yaqoob}}{{Dov{\v
  c}iak} et~al.}{2004}]{Dovciak04}
{Dov{\v c}iak} M.,  {Karas} V.,    {Yaqoob} T.,  2004, \apjs, 153, 205

\bibitem[\protect\citeauthoryear{{Fabian}, {Iwasawa}, {Reynolds} \&
  {Young}}{{Fabian} et~al.}{2000}]{Fabian00}
{Fabian} A.~C.,  {Iwasawa} K.,  {Reynolds} C.~S.,    {Young} A.~J.,  2000,
  \pasp, 112, 1145

\bibitem[\protect\citeauthoryear{{Fabian}, {Rees}, {Stella} \&
  {White}}{{Fabian} et~al.}{1989}]{Fabian89}
{Fabian} A.~C.,  {Rees} M.~J.,  {Stella} L.,    {White} N.~E.,  1989, \mnras,
  238, 729

\bibitem[\protect\citeauthoryear{{Fabian} \& {Ross}}{{Fabian} \&
  {Ross}}{2010}]{Fabian10}
{Fabian} A.~C.,  {Ross} R.~R.,  2010, \ssr, 157, 167

\bibitem[\protect\citeauthoryear{{Frank}, {King} \& {Lasota}}{{Frank}
  et~al.}{1987}]{Frank87}
{Frank} J.,  {King} A.~R.,    {Lasota} J.-P.,  1987, \aap, 178, 137

\bibitem[\protect\citeauthoryear{{Galloway}, {Psaltis}, {Muno} \&
  {Chakrabarty}}{{Galloway} et~al.}{2006}]{Galloway06}
{Galloway} D.~K.,  {Psaltis} D.,  {Muno} M.~P.,    {Chakrabarty} D.,  2006,
  \apj, 639, 1033

\bibitem[\protect\citeauthoryear{{George} \& {Fabian}}{{George} \&
  {Fabian}}{1991}]{George91}
{George} I.~M.,  {Fabian} A.~C.,  1991, \mnras, 249, 352

\bibitem[\protect\citeauthoryear{{Giacconi}, {Murray}, {Gursky}, {Kellogg},
  {Schreier}, {Matilsky}, {Koch} \& {Tananbaum}}{{Giacconi}
  et~al.}{1974}]{Giacconi74}
{Giacconi} R.,  {Murray} S.,  {Gursky} H.,  {Kellogg} E.,  {Schreier} E.,
  {Matilsky} T.,  {Koch} D.,    {Tananbaum} H.,  1974, \apjs, 27, 37

\bibitem[\protect\citeauthoryear{{Giles}, {Hill}, {Strohmayer} \&
  {Cummings}}{{Giles} et~al.}{2002}]{Giles02}
{Giles} A.~B.,  {Hill} K.~M.,  {Strohmayer} T.~E.,    {Cummings} N.,  2002,
  \apj, 568, 279

\bibitem[\protect\citeauthoryear{{Hiemstra}, {M{\'e}ndez}, {Done}, {D{\'{\i}}az
  Trigo}, {Altamirano} \& {Casella}}{{Hiemstra} et~al.}{2011}]{Hiemstra11}
{Hiemstra} B.,  {M{\'e}ndez} M.,  {Done} C.,  {D{\'{\i}}az Trigo} M.,
  {Altamirano} D.,    {Casella} P.,  2011, \mnras, 411, 137

\bibitem[\protect\citeauthoryear{{Hoffman}, {Lewin} \& {Doty}}{{Hoffman}
  et~al.}{1977}]{Hoffman77}
{Hoffman} J.~A.,  {Lewin} W.~H.~G.,    {Doty} J.,  1977, \apjl, 217, L23

\bibitem[\protect\citeauthoryear{{Iaria}, {D'A{\'{\i}}}, {di Salvo}, {Robba},
  {Riggio}, {Papitto} \& {Burderi}}{{Iaria} et~al.}{2009}]{Iaria09}
{Iaria} R.,  {D'A{\'{\i}}} A.,  {di Salvo} T.,  {Robba} N.~R.,  {Riggio} A.,
  {Papitto} A.,    {Burderi} L.,  2009, \aap, 505, 1143

\bibitem[\protect\citeauthoryear{{Kallman} \& {White}}{{Kallman} \&
  {White}}{1989}]{Kallman89}
{Kallman} T.,  {White} N.~E.,  1989, \apj, 341, 955

\bibitem[\protect\citeauthoryear{{Kuulkers}, {van der Klis}, {Oosterbroek},
  {Asai}, {Dotani}, {van Paradijs} \& {Lewin}}{{Kuulkers}
  et~al.}{1994}]{Kuulkers94}
{Kuulkers} E.,  {van der Klis} M.,  {Oosterbroek} T.,  {Asai} K.,  {Dotani} T.,
   {van Paradijs} J.,    {Lewin} W.~H.~G.,  1994, \aap, 289, 795

\bibitem[\protect\citeauthoryear{{Laor}}{{Laor}}{1991}]{Laor91}
{Laor} A.,  1991, \apj, 376, 90

\bibitem[\protect\citeauthoryear{{Li}, {Zimmerman}, {Narayan} \&
  {McClintock}}{{Li} et~al.}{2005}]{Li05}
{Li} L.-X.,  {Zimmerman} E.~R.,  {Narayan} R.,    {McClintock} J.~E.,  2005,
  \apjs, 157, 335

\bibitem[\protect\citeauthoryear{{Lin}, {Remillard} \& {Homan}}{{Lin}
  et~al.}{2007}]{Lin07}
{Lin} D.,  {Remillard} R.~A.,    {Homan} J.,  2007, \apj, 667, 1073

\bibitem[\protect\citeauthoryear{{Miller}}{{Miller}}{2007}]{Miller07}
{Miller} J.~M.,  2007, \araa, 45, 441

\bibitem[\protect\citeauthoryear{{Miller}, {Lamb} \& {Cook}}{{Miller}
  et~al.}{1998}]{Miller98}
{Miller} M.~C.,  {Lamb} F.~K.,    {Cook} G.~B.,  1998, \apj, 509, 793

\bibitem[\protect\citeauthoryear{{Mitsuda}, {Inoue}, {Koyama}, {Makishima},
  {Matsuoka}, {Ogawara}, {Suzuki}, {Tanaka}, {Shibazaki} \& {Hirano}}{{Mitsuda}
  et~al.}{1984}]{Mitsuda84}
{Mitsuda} K.,  {Inoue} H.,  {Koyama} K.,  {Makishima} K.,  {Matsuoka} M.,
  {Ogawara} Y.,  {Suzuki} K.,  {Tanaka} Y.,  {Shibazaki} N.,    {Hirano} T.,
  1984, \pasj, 36, 741

\bibitem[\protect\citeauthoryear{{Ng}, {D{\'{\i}}az Trigo}, {Cadolle Bel} \&
  {Migliari}}{{Ng} et~al.}{2010}]{Ng10}
{Ng} C.,  {D{\'{\i}}az Trigo} M.,  {Cadolle Bel} M.,    {Migliari} S.,  2010,
  \aap, 522, A96+

\bibitem[\protect\citeauthoryear{{Pandel}, {Kaaret} \& {Corbel}}{{Pandel}
  et~al.}{2008}]{Pandel08}
{Pandel} D.,  {Kaaret} P.,    {Corbel} S.,  2008, \apj, 688, 1288

\bibitem[\protect\citeauthoryear{{Piraino}, {Santangelo}, {di Salvo}, {Kaaret},
  {Horns}, {Iaria} \& {Burderi}}{{Piraino} et~al.}{2007}]{Piraino07}
{Piraino} S.,  {Santangelo} A.,  {di Salvo} T.,  {Kaaret} P.,  {Horns} D.,
  {Iaria} R.,    {Burderi} L.,  2007, \aap, 471, L17

\bibitem[\protect\citeauthoryear{{Reynolds} \& {Fabian}}{{Reynolds} \&
  {Fabian}}{2008}]{Reynolds08}
{Reynolds} C.~S.,  {Fabian} A.~C.,  2008, \apj, 675, 1048

\bibitem[\protect\citeauthoryear{{Ross} \& {Fabian}}{{Ross} \&
  {Fabian}}{1993}]{Ross93}
{Ross} R.~R.,  {Fabian} A.~C.,  1993, \mnras, 261, 74

\bibitem[\protect\citeauthoryear{{Ross} \& {Fabian}}{{Ross} \&
  {Fabian}}{2005}]{Ross05}
{Ross} R.~R.,  {Fabian} A.~C.,  2005, \mnras, 358, 211

\bibitem[\protect\citeauthoryear{{Ross}, {Fabian} \& {Young}}{{Ross}
  et~al.}{1999}]{Ross99}
{Ross} R.~R.,  {Fabian} A.~C.,    {Young} A.~J.,  1999, \mnras, 306, 461

\bibitem[\protect\citeauthoryear{{Sanna}, {M{\'e}ndez}, {Belloni} \&
  {Altamirano}}{{Sanna} et~al.}{2012}]{Sanna12}
{Sanna} A.,  {M{\'e}ndez} M.,  {Belloni} T.,    {Altamirano} D.,  2012, \mnras,
  424, 2936

\bibitem[\protect\citeauthoryear{{Shakura} \& {Sunyaev}}{{Shakura} \&
  {Sunyaev}}{1973}]{Shakura73}
{Shakura} N.~I.,  {Sunyaev} R.~A.,  1973, \aap, 24, 337

\bibitem[\protect\citeauthoryear{{Shih}, {Bird}, {Charles}, {Cornelisse} \&
  {Tiramani}}{{Shih} et~al.}{2005}]{Shih05}
{Shih} I.~C.,  {Bird} A.~J.,  {Charles} P.~A.,  {Cornelisse} R.,    {Tiramani}
  D.,  2005, \mnras, 361, 602

\bibitem[\protect\citeauthoryear{{Strohmayer} \& {Markwardt}}{{Strohmayer} \&
  {Markwardt}}{2002}]{Strohmayer02}
{Strohmayer} T.~E.,  {Markwardt} C.~B.,  2002, \apj, 577, 337

\bibitem[\protect\citeauthoryear{{Str{\"u}der}, {Aschenbach}, {Br{\"a}uninger},
  {Drolshagen}, {Englhauser}, {Hartmann}, {Hartner}, {Holl}, {Kemmer},
  {Meidinger}, {St{\"u}big} \& {Tr{\"u}mper}}{{Str{\"u}der}
  et~al.}{2001}]{Strueder01}
{Str{\"u}der} L.,  {Aschenbach} B.,  {Br{\"a}uninger} H.,  {Drolshagen} G.,
  {Englhauser} J.,  {Hartmann} R.,  {Hartner} G.,  {Holl} P.,  {Kemmer} J.,
  {Meidinger} N.,  {St{\"u}big} M.,    {Tr{\"u}mper} J.,  2001, \aap, 365, L18

\bibitem[\protect\citeauthoryear{{Svoboda}, {Dovciak}, {Goosmann} \&
  {Karas}}{{Svoboda} et~al.}{2009}]{Svoboda09}
{Svoboda} J.,  {Dovciak} M.,  {Goosmann} R.~W.,    {Karas} V.,  2009, ArXiv
  e-prints

\bibitem[\protect\citeauthoryear{{Tanaka}, {Nandra}, {Fabian}, {Inoue},
  {Otani}, {Dotani}, {Hayashida}, {Iwasawa}, {Kii}, {Kunieda}, {Makino} \&
  {Matsuoka}}{{Tanaka} et~al.}{1995}]{Tanaka95}
{Tanaka} Y.,  {Nandra} K.,  {Fabian} A.~C.,  {Inoue} H.,  {Otani} C.,  {Dotani}
  T.,  {Hayashida} K.,  {Iwasawa} K.,  {Kii} T.,  {Kunieda} H.,  {Makino} F.,
   {Matsuoka} M.,  1995, \nat, 375, 659

\bibitem[\protect\citeauthoryear{{Titarchuk}, {Kazanas} \&
  {Becker}}{{Titarchuk} et~al.}{2003}]{Titarchuk03}
{Titarchuk} L.,  {Kazanas} D.,    {Becker} P.~A.,  2003, \apj, 598, 411

\bibitem[\protect\citeauthoryear{{van der Klis}}{{van der Klis}}{2006}]{Klis06}
{van der Klis} M.,  2006, {Rapid X-ray Variability}.
pp 39--112

\bibitem[\protect\citeauthoryear{{van Paradijs}, {van der Klis}, {van
  Amerongen}, {Pedersen}, {Smale}, {Mukai}, {Schoembs}, {Haefner}, {Pfeiffer}
  \& {Lewin}}{{van Paradijs} et~al.}{1990}]{Paradijs90}
{van Paradijs} J.,  {van der Klis} M.,  {van Amerongen} S.,  {Pedersen} H.,
  {Smale} A.~P.,  {Mukai} K.,  {Schoembs} R.,  {Haefner} R.,  {Pfeiffer} M.,
  {Lewin} W.~H.~G.,  1990, \aap, 234, 181

\bibitem[\protect\citeauthoryear{{Verner}, {Ferland}, {Korista} \&
  {Yakovlev}}{{Verner} et~al.}{1996}]{Verner96}
{Verner} D.~A.,  {Ferland} G.~J.,  {Korista} K.~T.,    {Yakovlev} D.~G.,  1996,
  \apj, 465, 487

\bibitem[\protect\citeauthoryear{{Wijnands}, {van der Klis}, {van Paradijs},
  {Lewin}, {Lamb}, {Vaughan} \& {Kuulkers}}{{Wijnands}
  et~al.}{1997}]{Wijnands97}
{Wijnands} R.~A.~D.,  {van der Klis} M.,  {van Paradijs} J.,  {Lewin} W.~H.~G.,
   {Lamb} F.~K.,  {Vaughan} B.,    {Kuulkers} E.,  1997, \apjl, 479, L141

\bibitem[\protect\citeauthoryear{{Willmore}, {Mason}, {Sanford}, {Hawkins},
  {Murdin}, {Penston} \& {Penston}}{{Willmore} et~al.}{1974}]{Willmore74}
{Willmore} A.~P.,  {Mason} K.~O.,  {Sanford} P.~W.,  {Hawkins} F.~J.,  {Murdin}
  P.,  {Penston} M.~V.,    {Penston} M.~J.,  1974, \mnras, 169, 7

\bibitem[\protect\citeauthoryear{{Wilms}, {Allen} \& {McCray}}{{Wilms}
  et~al.}{2000}]{Wilms00}
{Wilms} J.,  {Allen} A.,    {McCray} R.,  2000, \apj, 542, 914

\bibitem[\protect\citeauthoryear{{Zdziarski}, {Johnson} \&
  {Magdziarz}}{{Zdziarski} et~al.}{1996}]{Zdziarski96}
{Zdziarski} A.~A.,  {Johnson} W.~N.,    {Magdziarz} P.,  1996, \mnras, 283, 193

\bibitem[\protect\citeauthoryear{{Zhang}, {M{\'e}ndez} \& {Altamirano}}{{Zhang}
  et~al.}{2011}]{Zhang11}
{Zhang} G.,  {M{\'e}ndez} M.,    {Altamirano} D.,  2011, \mnras, 413, 1913

\bibitem[\protect\citeauthoryear{{Zhang}, {Lapidus}, {Swank}, {White} \&
  {Titarchuk}}{{Zhang} et~al.}{1997}]{Zhang97}
{Zhang} W.,  {Lapidus} I.,  {Swank} J.~H.,  {White} N.~E.,    {Titarchuk} L.,
  1997, \iaucirc, 6541, 1

\bibitem[\protect\citeauthoryear{{{\.Z}ycki}, {Done} \& {Smith}}{{{\.Z}ycki}
  et~al.}{1999}]{Zycki89}
{{\.Z}ycki} P.~T.,  {Done} C.,    {Smith} D.~A.,  1999, \mnras, 309, 561

\end{thebibliography}
%
%
%

%
\bsp
\label{lastpage}

\end{document}